\documentclass[onecolumn]{article}
\usepackage[dvips]{graphicx}

\usepackage{cuted}
\usepackage{lipsum, color}

\usepackage[noadjust]{cite}
\usepackage{filecontents}

\usepackage{bm}

\usepackage[margin=1in]{geometry}
\usepackage{amsmath}
\usepackage{amssymb}
\usepackage{algorithm}
\usepackage{algorithmicx}
\usepackage[noend]{algpseudocode}
\newcommand{\argmax}{\operatornamewithlimits{argmax}}

\usepackage{epstopdf}
  
\newcommand{\rom}[1]{%
  \textup{\uppercase\expandafter{\romannumeral#1}}%
}
\usepackage{authblk}
\usepackage{amsmath,amssymb,verbatim,graphicx,multirow,amsfonts,textcomp}
\usepackage{times}
\usepackage{epsfig}
\usepackage{graphicx}
\usepackage{amsmath}
\usepackage{amssymb}
\usepackage{rotating}
\usepackage{url}
\usepackage{verbatim,subfigure,multirow}
\usepackage{array, blindtext}
\usepackage{bbm}
\usepackage{setspace}
\usepackage{float}
\usepackage{multirow}
\restylefloat{table}
\usepackage[table]{xcolor}
\usepackage{xspace}
\makeatletter
\DeclareRobustCommand\onedot{\futurelet\@let@token\@onedot}
\def\@onedot{\ifx\@let@token.\else.\null\fi\xspace}
\def\eg{\emph{e.g}\onedot} 
\def\ie{\emph{i.e}\onedot}

\def\etal{\emph{et al}\onedot}

\newcommand{\blue}[1]{\textcolor{blue}{#1}}

\newcommand{\x}{\mathbf{x}}
\newcommand{\y}{\mathbf{y}}
\newcommand{\h}{\mathbf{h}}

\newcommand{\X}{\mathbf{X}}

\newcommand{\calX}{\mathcal{X}}
\newcommand{\calY}{\mathcal{Y}}

\newcommand{\T}{\mathcal{T}}
\newcommand{\W}{\mathcal{W}}

\author[a]{Nazli Farajidavar} 
\author[b]{Sefki Kolozali} 
\author[c]{Payam Barnaghi}

\affil[a]{Institute for Biomedical Engineering, University of Oxford, UK}
\affil[b]{MRC-PHE Centre for Environment \& Health, Analytical \& Environmental Sciences, Faculty of Life Sciences \& Medicine, King's College London, UK}
\affil[c]{Institute for communication systems (ICS), University of Surrey, UK}

\title{\centering A Deep Multi-View Learning Framework for City Event Extraction from Twitter Data Streams}
\date{}

\begin{document}
\maketitle
\begin{abstract}
Cities have been a thriving place for citizens over the centuries due to
their complex infrastructure. The emergence of the Cyber-Physical-Social Systems (CPSS)
and context-aware technologies boost a growing interest in analysing, extracting 
and eventually understanding city events which subsequently can be utilised to 
leverage the citizen observations of their cities. In this paper, we 
investigate the feasibility of using Twitter textual streams for extracting city events.
We propose a hierarchical multi-view deep 
learning approach to contextualise citizen observations of various city systems
and services such as traffic, public transport, weather, sociocultural activities 
and public safety as a source of city events. Our goal has been to build a flexible 
architecture that can learn representations useful for tasks, thus avoiding excessive 
task-specific feature engineering. We apply our approach on a real-world dataset 
consisting of event reports and tweets collected by~\cite{Pramod-ACM-2015} over four 
months from San Francisco Bay Area dataset and additional datasets collected from Greater London. 
The results of our evaluations show that our proposed solution outperforms the existing models and 
can be used for extracting city related events with an averaged accuracy of $81\%$ over all classes.
To further evaluate the impact of our Twitter event extraction model, we have used two sources of authorised reports through collecting road traffic disruptions data from Transport for London API, and parsing the Time Out London website for sociocultural events. The analysis showed that 49.5\% of the Twitter traffic comments are reported approximately five hours prior to the authorities official records. Moreover, we discovered that amongst the scheduled sociocultural event topics; tweets reporting transportation, cultural and social events are 31.75\% more likely to influence the distribution of the Twitter comments than sport, weather and crime topics.
\end{abstract}

\section{Introduction}
\label{sec:intro}

\begin{figure}
\centering     
\subfigure[Figure A]{\label{fig:cityDepartments}\includegraphics[width=0.6\columnwidth]{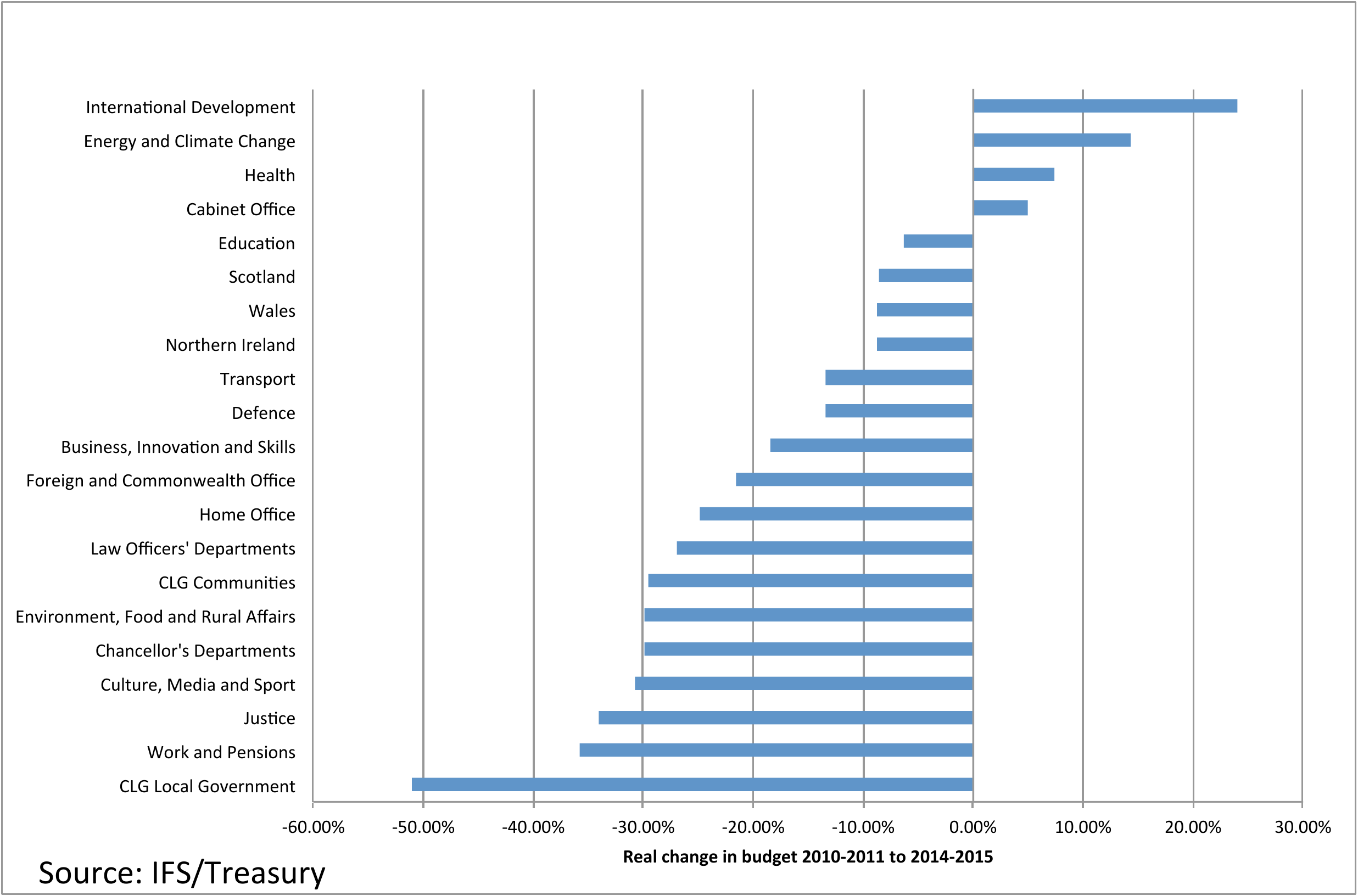}}
\subfigure[Figure B]{\label{fig:twitter-sample}\includegraphics[width=0.3\textwidth]{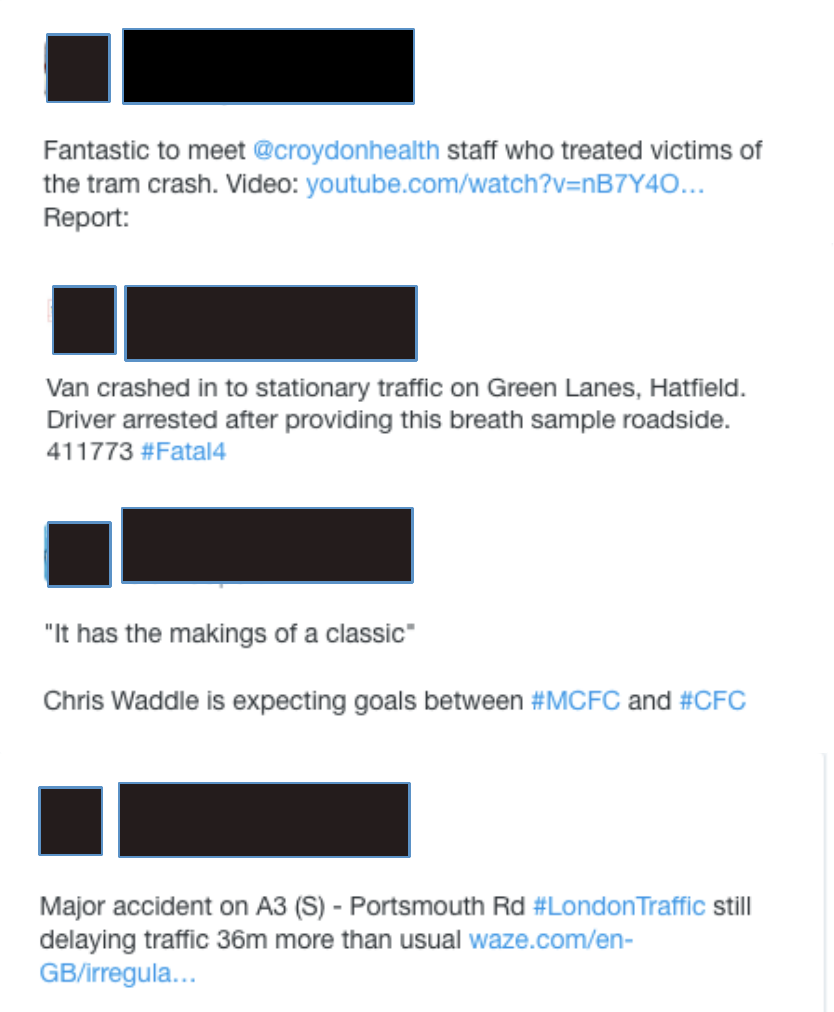}}
\caption{A: Common city departments [\textcopyright~\cite{SmartCities2011}], B: Tweets reporting various concerns about a city spanning power supply, water 		quality, traffic jams, and public transport delays (\textcopyright~\cite{Pramod-ACM-2015}).}
\end{figure}

Recent advances in ubiquitous computing and context-aware technologies have boosted the interest in smart city framework designs. 
These frameworks endeavour to provide authorities and citizens with real-time information and assistance in the decision-making and resource allocation processes. 
Meantime, the departmental structure of a city can be very complex, and its management continues to be strained by various factors, such as dynamic nature of their services, population growth and continuously shrinking pool of available financial resources. 
Figure~\ref{fig:cityDepartments} illustrates an evidence of some of the common departments that provide public support and management for London and their budget re-allocations within the past two years.~\footnote{https://www.gov.uk/government/publications/public-expenditure-statistical-analyses-2015}

Some of the services offered by these departments are dynamic, \eg,
transportation services and their behaviour may vary in response to social and cultural events, 
accidents, and weather conditions. 
In this sense, understanding events occurring in cities is of great contemporary 
interest~\cite{Naphade:2011:SCI:1999160.1999174,Lindsey2010,kehoe:Cosgrove:2011} 
to city authorities to enhance their management and to optimise operations and interactions 
among various city departments and services. A possible way to do this is through getting continues feedback and event reports from citizens, who are the front-end users of these services.

Meanwhile, the emergence of social networks, such as Twitter\footnote{http://twitter.com/}, Facebook\footnote{http://facebook.com/} and Instagram\footnote{http://www.instagram.com}, 
offers enormous information that can be exploited for citizen sensing. This could be used to notify citizens as well as authorities regarding the events occurring in smart urban spaces (Figure~\ref{fig:twitter-sample} depicts samples of real-world city
events reported directly by citizens on social media).
However, the citizen sensing~\cite{Sheth2009,Burke06participatorysensing}
component that can provide complementary or corroborative information is often
ignored in state-of-the-art analytics for smart cities~\cite{Filipponi:2010}.

In this article we propose a hybrid pipeline for real-time sensing in cities through utilisation of complementary dynamic data sources, namely Twitter, London Road disruption reports from traffic sensors; and Time Out London. The proposed data processing pipeline involves data wrappers, a novel Natural Language Processing (NLP) component based on multi-view learning, and multi-sensor correlation analysis. We presented a priliminary version of this pipeline in~\cite{DBLP:conf/wf-iot/FarajiDavarKB16}. And in this article we will further focus on the detailed theoretical design aspects of the model and include extended experiments to showcase its performance.
The multi-view learning component combines the output of a 
Convolutional Neural Network (CNN) learning with a name entity event extraction to enable a near real-time city-related event extraction from short informal text corpus of Twitter.
Developing a scalable automatic city event annotation system, 
we show that our proposed solution achieves performance boost compared to the state-of-the-art approaches~\cite{Pramod-ACM-2015,Ritter:2012:ODE:2339530.2339704}. Up to our knowledge, this is the first time that a multi-view deep learning algorithm
has been proposed in the context of city event extraction. Subsequently, we conducted a similarity analysis on the processed data from social media, road sensors, and Web of Data, and discover the associations between incidents in near real-time.
 The research contributions are four-fold in high-level and can be summarised as follows: i) Automated real-time data collection wrappers for Twitter and city sensors; ii) A near real-time NLP component for classifying Twitter data; iii) A correlation analysis for detecting the dependencies between Twitter stream and city sensors and web driven data records; iv) A web interface for displaying and visualising the city’s event highlights. The fine-grained contributions of the proposed NLP component are as follows: ii-i) real-time multi-label event extraction from Twitter, ii-ii) a novel multi-view deep learning formulation for event extraction based on graphical models, ii-iii) late  classification results fusion for an enhanced event location extraction from tweets.

The paper is organised as follows. Section~\ref{sec:related-work} describes the benchmark task of interest - Tweet annotation -  
where we discuss related works. In Section~\ref{sec:method}, 
we describe the proposed multi-view pipeline. Section~\ref{sec:exp} details our 
experimental setup and discusses the evaluation results. 
Finally, in Section~\ref{sec:conclusion} we derive a conclusion for our work and provide future directions.

\section{Related work} 
\label{sec:related-work}
Typically, a city has many departments such as public safety, urban planning, energy, water, transportation, social programs, and education~\cite{Blissent:2010,Blissent:2013}. 
The live updates on the performance and quality of services offered by these departments are
important for city authorities to leverage the management of city resources and for citizens to make more informed decisions using the city services and to interact better with surrounding environment. 
Meanwhile, social media networks, such as Twitter offer a near real-time communication platforms which can be utilised to facilitate this purpose. 
Such information can complement sensor data and textual reports collected from conventional sources or city departments, and it can help to enhance the public services. 
For example, sensors deployed on a road may report reduced speed of vehicles which can 
be explained by the procession obstructing traffic that is reported on social media.

The design of such platform which utilises the social media as a source for public sensing in city-related event extraction context, needs to address the following research question: How to extract city infrastructure
related events from Twitter? How to exploit event and location knowledge-bases
for event extraction? And finally how accurately these Twitter extracted events are matching the reality of city events?

The studies such as~\cite{Moraru:2012,Pramod-ACM-2015} assumed the presence of event data sources such
as sensory data (\eg, loop detectors) and formal report of events 
(\eg, eventful\footnote{http://eventful.com/}) in a city. 
While utilisation of such a formal data source can serve as a reliable source for training an automated event extraction system, such resources may not be available with short latency or even not exist at all in many cities. Therefore, we need the alternative and complementary data sources for training such model for different cities.

Event extraction from textual corpus, can be categorised  into two groups according to the structure of the text; formal corpus vs informal. Where the former
refers to the grammatical text such as news documents and the later addresses 
the user-generated content with no overt structure that might contain a lot of 
slang and non-standard abbreviations and notations (as it is the case in data obtained from Twitter).

In formal text analysis domain, Liu \etal~\cite{DBLP:conf/ideal/LiuLXCY08} proposed 
to alleviate information overload in daily news by extracting key entity and 
significant event of news documents.
A bipartite graph was induced in~\cite{Asratian:1998:BGA:294028}, based on the entities 
and their associations to documents using mutual 
reinforcement principle capturing salient entities and the documents with salient entities used 
to rank the news events. Extraction of local events from blog entries carried
out by~\cite{conf/airs/OkamotoK09}.
Use of lightweight patterns to extract global crisis events from news
text presented in \cite{conf/nldb/TanevPA08}.
Event extraction in the context of detecting infectious disease outbreak was achieved 
by~\cite{Grishman:2002:REE:1289189.1289229} where the event schema consisted of date range, geo-location, disease name, organism type
and number affected by the disease, and the organism survival information. The event extraction then obtained by finite-state pattern matching on the tokenized input text. More recently, adding convolutional layers to the neural network language model of Bengio \etal~\cite{Bengio:2003:NPL:944919.944966}, Collorbert \etal~\cite{Collobert:2011:NLP:1953048.2078186} developed their convolutional neural network model that shared representations across the tasks of language modelling, part of speech tagging, chuncking, named entity recognition, semantic role labelling, and syntactic parsing. Although the proposed model was not specifically designed for event extraction, its performance surpassed the state of the art methods on majority of the language modelling tasks.

Event extraction from informal text (which is our main focus in this paper 
due to the informal nature of Twitter textual content) 
is also addressed in literature~\cite{DBLP:conf/icwsm/BeckerNG11,Ritter:2012:ODE:2339530.2339704,Pramod-ACM-2015}.
In~\cite{DBLP:conf/icwsm/BeckerNG11}, the authors used temporal (volume
changes), social (replies, broadcast), topical (coherence of clusters), and Twitter-centric
(multi-word hashtags) features to train a classifier that performed better than the baseline.
Ritter \etal~\cite{Ritter:2012:ODE:2339530.2339704}, solved the task in an unsupervised manner by building a calendar of significant
events such as sports, concert, protests, politics, TV, and religion. 
Their approach utilised the Latent Drichlet Allocations (LDA) method to model each entity 
in terms of a mixture of event types and each event type in terms of a mixture of entities. 
Recent stdudies in \cite{wang2012automatic, PCS/Zhou16} utilised the LDA for hit and run crimes and traffic related event extraction, respectivlely. And in~\cite{Marquez:2008:SRL:1403157.1403158}, the authors used the latent topic model for semantic role labeling task in Twitter data. 
A generalised linear regression model also used to capture the association 
between topics and crimes from a training dataset.
Lampos and Cristianini~\cite{Lampos:2012:NES:2337542.2337557} proposed to 
use an optimised feature selection approach with a
regressor to estimate the intensity of environmental and epidemiological events based on event markers.

Considering the same assumption as of~\cite{Moraru:2012}, Anantharam \etal~\cite{Pramod-ACM-2015} developed an automatic data annotation unit to obtain ground truth by using officially reported traffic events~\footnote{http://511.org} and location~\footnote{https://www.openstreetmap.org/} knowledge-bases. The authors then used this annotated data to train a CRF-based event extraction model to capture long-term word dependencies for Twitter analysis. While their proposed approach for the preparation of the ground-truth data has shown a good word-tagging performance, the proposed CRF-based event extraction had some limitations. The model was designed to only extract traffic events. Precisely speaking, since the automatic annotation unit was trained with the officially reported ground-truth traffic events of a limited time period, the model performed poorly in the prediction of future incidents specifically reported by new users.  Besides, although the location terms have been extracted, they were not utilised to associate locations with extracted events. Instead, the authors assumed that the tweet's geo-location tag (the location where the users tweet the events) can serve as the event locations, which is not always valid.

In multi-view learning literature, Chen \etal \cite{conf/nips/ChenZX10} developed a statistical framework that learns a predictive subspace shared by multiple views based on a generic multi-view latent space Markov network. Kumar \etal~\cite{daume11spectral} co-trained unsupervised learning models and proposed a spectral clustering algorithm for multi-view data. Quadrianto and Lampert~\cite{kernel-based-Learning2013} studied the metric learning problem in cross-media retrieval tasks with the aim to learn metrics with which the original multi-view higher dimensional features can be projected into a shared feature space, so that the Euclidean distance in this space is meaningful not only within a single view, but also among different views. In our multi-view learning model we used the Restricted Boltzmann Machines formulation to be consistent with the rest of the neural network architecture of CNNs. 
  
While in all proposed platforms for event extraction from Twitter, the main focus had been on training an NLP model using tweet's informal text corpus, the human intelligence learning model does not work as such. As human, we initially learn the semantic meaning of the words in a language. We then, been taught on the synthetic structure (Grammar rules) of the sentence at the school by means of formal corpus (\ie books). Analogously, NLP approaches which are jointly attempt to accomplish the PoS and NER tagging using the informal Twitter corpus will not acquire the potential of being extended to future data due to their intrinsic limitation.  
Taking this into consideration, we have proposed an NLP framework for informal text classification which is not only applicable to future data but also addresses the the limitations of the other state of the art approaches. We utilised a CRF-based  Name Entity Recognition (NER) model 
of~\cite{Pramod-ACM-2015} and extending it beyond traffic event extraction, 
we have proposed a multi-view learning pipeline which fuses the CRF output with 
the part of speech (POS) tags extracted from the Convolutional Neural 
Network (CNN)~\cite{Collobert:2011:NLP:1953048.2078186} model, 
for leveraging the city event extraction. 
  
Utilising a CNN model which is trained on formal texts for PoS tagging of tweet words is plausible, since the underlying syntactic role of words in a language are still valid even in informal texts such as Twitter corpora despite their variation in sentence grammatical structures.
In terms of CRF training, unlike Anantharam~\cite{Pramod-ACM-2015} \etal's model, our proposed model is trained on more generic categorical data and is capable of detecting a wider categorical range of city events.
This allows the model to better generalise to future events and incidents. 
While various neural network architectures~\cite{C14-1008,Glorot+al-ICML-2011,S14-2033} have 
been proposed in literature and their performance are investigated for Twitter 
sentiment classification, to the best of our knowledge, this is the first time that the CNN text 
analysis is utilised for city even extraction from informal text and its result is integrated  
with a CRF NER tagger in a deep multi-view learning framework to obtain an enhanced 
sentence-level inference and event extraction. To further validate the verity of the extracted events, we have parsed data from London Traffic API and TimeOut London sociocultural resources and evaluated the veracity of twitter extracted events through a graph-based similarity analysis.

\section{Methodology}
\label{sec:method}
\begin{figure}[h!]\centering
\includegraphics[width=.7\linewidth]{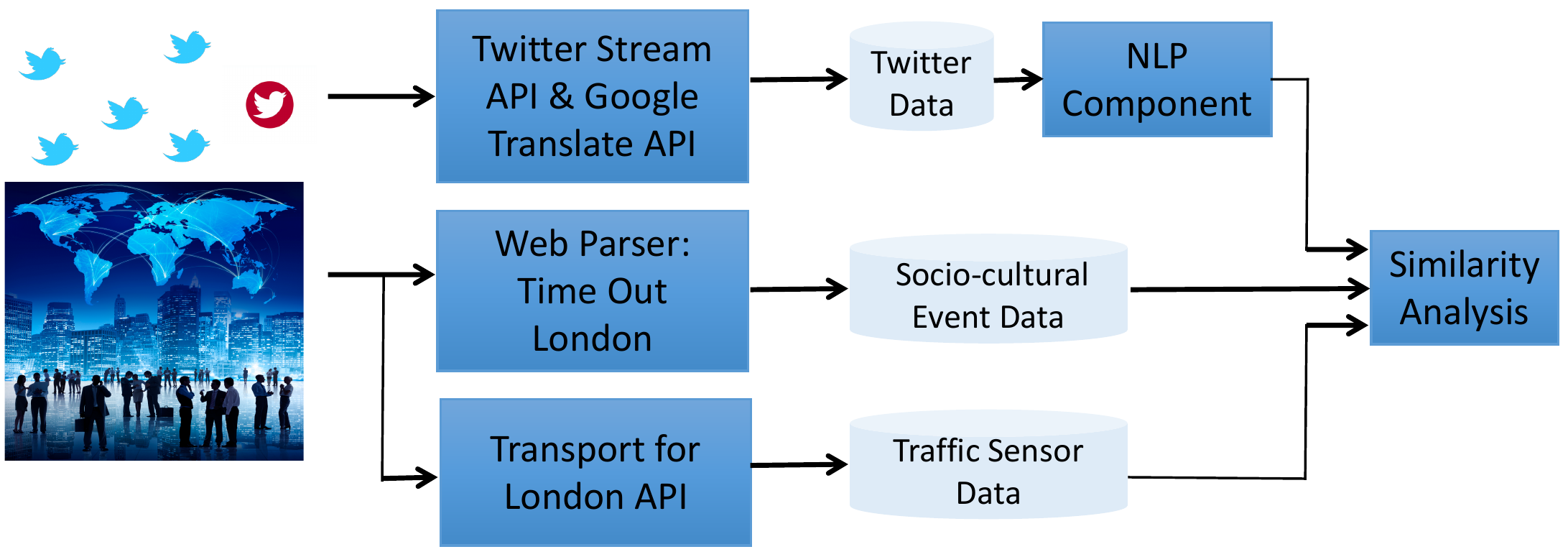}
\caption{The proposed hybrid pipeline}
\label{fig:SM_pipeline}
\end{figure}

Our proposed hybrid approach is based on undirected graphical models. Figure~\ref{fig:SM_pipeline} depicts the diagram of the proposed hybrid approach. We developed three data wrappers to collect data from the city; Twitter stream API~\footnote{https://dev.twitter.com/streaming}, Transport for London API~\footnote{http://data.tfl.gov.uk/tfl}, and Time Out London~\footnote{http://www.timeout.com/london} parser. Furthermore, we developed a data processing component that involves of two main parts: i) Natural Language Processing (NLP) on Twitter data streams and ii) similarity analysis on Twitter, road sensor data, and scheduled events collected from Time Out London website. We used the Google translate API to automatically detect the source language on non-English tweets and translate them into English to facilitate the text analysis step.

\subsection{Twitter NLP Component}
\label{sec:NLP_Comp}

\begin{figure}[h!]\centering
\includegraphics[width=.7\linewidth]{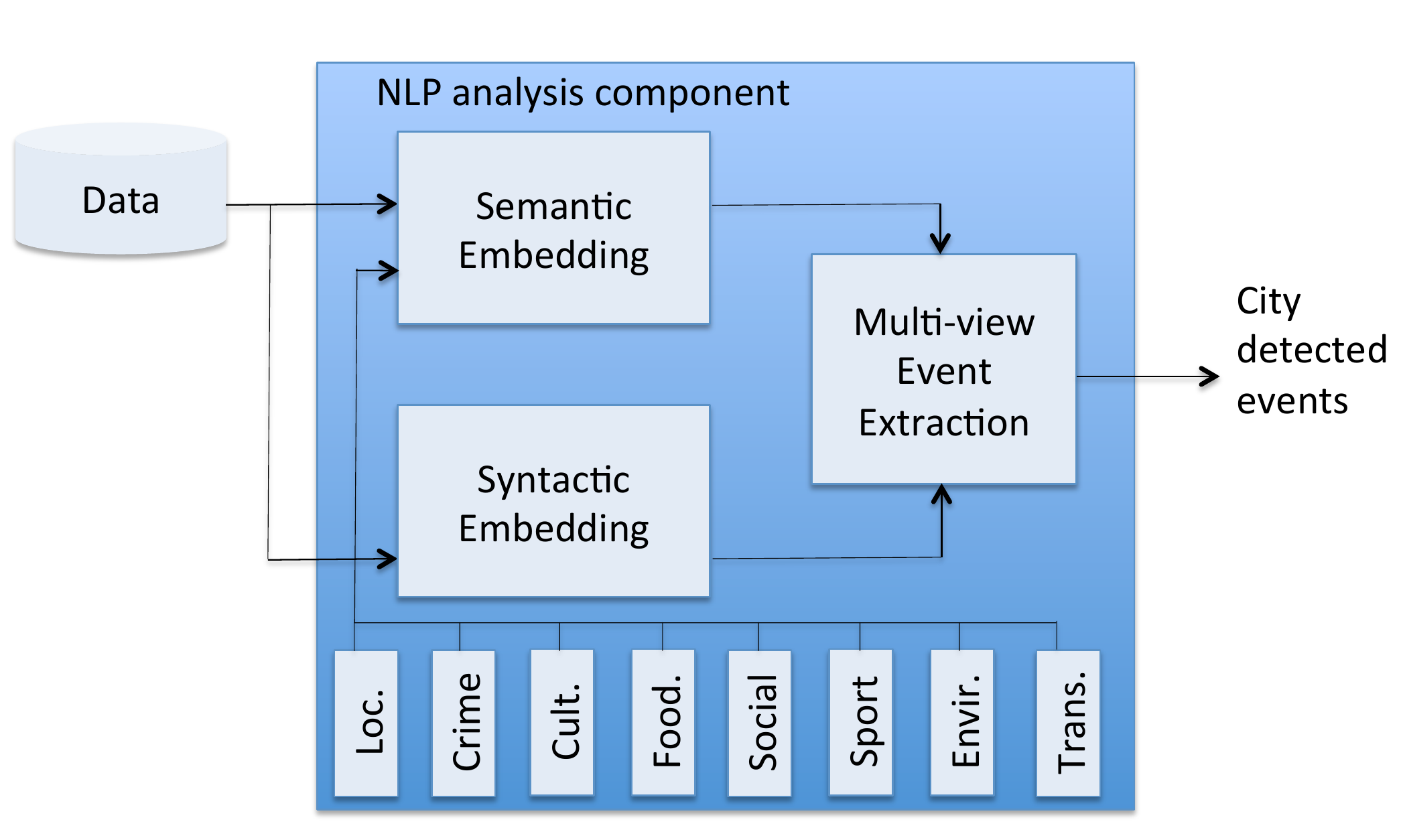}
\caption{NLP component: detailed event detection pipeline. Note that the semantic embedding view is modelled with a CRF and the syntactic embedding view is modelled through CNN.}
\label{fig:NLP_pipeline}
\end{figure}

Figure~\ref{fig:NLP_pipeline} shows the data processing units of the proposed NLP component which is composed of three sub-components: a semantic embedding subspace learning, a syntactic embedding subspace learning, and a multi-view event extraction.

Given a tweet text represented by $\x_i=words(Tweet_i)$, we are interested in associating it with one or multiple 
city-related event classes from the events set: $C$ = \{\textit{TransportationEvent,
WeatherEvent, CulturalEvent, SocialEvent, SportEvent, FoodEvent, CriminalEvent}\} along with a \textit{Location} tag.
To assign event tags to tweets, we have assumed that each tweet contains only one sentence. Considering the 140 character limit of a tweet, this assumption sounded plausible. We then decomposed sentences into semantic and syntactic embeddings where the former deals with the meaning of the words in the sentence and the later addresses its grammar structure.

The fusion of these embeddings have been used to provide an explicit insight to the meaning of sentences to facilitate their classification. This fusion can be formulated as a multi-view learning task where each embedding contributes to a distinct view of the same training data. Although baseline methods such as one proposed by~\cite{Pramod-ACM-2015} had shown an acceptable performance on time and location dependent annotation tasks, they will not generalise well to annotation task of varying locations and times. To address these generalisation issue, we have estimated the  semantic and syntactic embedding matrices off line and independently, using more comprehensive data. 

Inspired by human cognitive ability, we believe that a Part of Speech (PoS) word tagging approach which has been trained on encyclopedia corpus can help in extracting a more realistic syntactic embedding of the tweet. This in practice can resemble human’s general grammar knowledge. Doing though, we have adopted the CNN graphical model (CNN) proposed in~\cite{Collobert:2011:NLP:1953048.2078186} which had been trained on entire English Wikipedia. 

To align the formulation of the semantic embedding extraction with the CNN based syntactic embedding, and to capture the long term dependencies in name phrases, we have chosen the Conditional Random Field (CRF) formulation of undirected graphical models for Name Entity Recognition (NER). 
To do so, we have used phrases, short reports and location terms extracted from official websites and authority reports (listed in Table~\ref{tab:event-classes}) to built class conditional corpora. These conditional corpora are then used to train CRF models for name entity recognition.  

  \begin{table}[t]
  \begin{center}
  \begin{tabular}{|c|l|}
  \hline
  \textit{Event Class}  & Vocabulary Source \\ \hline
  Crime &\url{http://www.shouselaw.com/crimes-a-z.html}\\ \hline
  Cultural event &\url{http://en.wikipedia.org/wiki/Category:Cultural_events} \\ \hline
  Food & \url{http://www.foodterms.com/encyclopedia} \\ \hline
  Location &\url{https://www.openstreetmap.org}\\ \hline 
  Social event &\url{http://en.wikipedia.org/wiki/Category:Social_events}\\ \hline
    Sport & sport dictionaries of \cite{Ritter:2012:ODE:2339530.2339704}\\ \hline
  Weather &\url{http://www.erh.noaa.gov/er/box/glossary.htm}\\ \hline
  Transportation & \url{http://511.org} \\ \hline
  \end{tabular}
  \end{center}
  \caption{City-related event classes and their corresponding sample tweets}
  \label{tab:event-classes}
  \end{table}

To fuse the information gained from these two embeddings, we have proposed a multi-view learning approach. In order to be consistent with the rest of the architecture, we have chosen a supervised learning undirected graphical model, Restricted Boltzmann Machine (RBM). This formulation in practice uses the obtained tags of the two previous embeddings for mutually validating and scoring them for a final sentence-level inference~\footnote{Note that retraining the last supervised learning layer of a deep architecture is a common practice in deep learning}.
  
  An example of sentence level inference is 
  in the case of tweets such as ``seeing someone being given a parking ticket'' where individual
  words ``parking'' and ``ticket'' can belong to classes \textit{Transportation} and 
  \textit{Cultural} events respectively while considering these words' grammar roles can resolve
  this confusion.
  
  The output of the system can be represented as $<\hat{e_{type}}, \hat{e_{loc}}, \hat{t_{loc}}, \hat{t_{time}}, \hat{e_{impact}}>$ where $<\hat{e_{type}}, \hat{e_{loc}}$ are representing the event type and location, extracted from the proposed NLP analysis framework, $\hat{t_{loc}}, \hat{t_{time}}$ are tweet’s geo-location and time of report (meta data obtained from Twitter Streaming API) and finally $\hat{e_{impact}}>$ denotes the event impact. The event annotation impact score is calculated as the product of event severity and event likelihood scores as in~\cite{StatisticalMethods4Immunogenicity2015}.

\subsubsection{\bf{CRF Name Entity Tagging}}
\label{sec:CRF Tweets Annotation}
  The CRF is an undirected graphcal model~\cite{Koller+Friedman:09} 
  containing nodes that correspond to the
  set: $words(Tweet_i) \bigcup \T$ where  
  $ words(Tweet_i) =\x_i= \{w_1 , w_2 , ..., w_M\}$ and $\T = Tag_{set}$. 
  The model defines factors between (a) neighbouring tags
  $(tag_j , tag_{j+1} )$ and (b) tags and words $(tag_j , word_j)$ 
  in a sequence where $tag_j \in \T$ and
  $w_j \in words(Tweet_i)$. The factor function maps all possible values of inputs
  variable combinations to Real numbers (also known as potential for the input variable combination) and can be formulated as,
  $V \rightarrow R$ where $V \subset
  words(Tweet_i) \bigcup \T$, \eg, $\Phi(tag_j, tag_{j+1})$ captures the number of times $tag_j$ appears
  before $tag_{j+1}$ in a text. Concretely, if $tag_j$ is 
  B-Location representing the beginning location term, and $tag_{j+1}$ is I-Location, representing the intermediate location term, $\Phi(B-
  Location, I-Location)$ maps to the number of times this sequence appears in the corpus which may not be a normalised value. 
  The factors $\Phi(tag_j , w_j)$ for each word (where the $w_j$ is always observed)
  captures the number of times the word $w_j$ was labelled with the $tag_j$.
  Let's assume $w_j$ is a word \eg ``Piccadily'' and $tag_j$ is B-Location, then $\Phi(B-Location, Piccadily)$ 
  captures the number of times the word ``Piccadily'' was labelled with the tag B-Location in the
  corpus. 

  More specifically, if there are $|words(Tweet_i)|$ words in
  a tweet sequence, we need $(|words(Tweet_i)| - 1)$ factors to define relations between neighbouring tags and
  $|words(Tweet_i)|$ factors to define the relation between tags and words. 
  Finding the most likely tag assignment to a word in a tweet can be formalised as maximising the 
  probability $P(tags|words(Tweet_i))$ as shown in Table~\ref{tab:CRF-architecture}~(a). 

  Essentially, the tag assignment resulting in the highest probability
  score is chosen as the final tag assignment for all the words. Even though the
  model captures the relation between adjacent tags, tag assignment is done based on the
  global maximum \ie, tags that result in highest overall score are assigned to all the
  words. Such a global assignment of tags naturally captures long distance dependencies
  in text. 

  The location and event tagging module uses the linear chain CRF model
  presented in Table~\ref{tab:CRF-architecture}~(b) which is implemented in 
  LingPipe~\cite{LingPipe}. 
  In a linear chain CRF model, each tag type and its positions in a corpus are extracted using a 
  feature extractor function $f$ which takes
  position and the tags as input. The first word in the sequence will have \textit{``null''} 
  as the previous tag. 
  For the rest of the words in the input sequence, the feature function is invoked
  with all possible tags $(1,.., i, ..., M)$. $\beta(M)$ are the coefficient vectors 
  learned for each output tag in the tag set $\T$ where $M$ is the number of tags from the corpus. 
  The corresponding scores for tag assignment given words is provided as a 
  regression model and is not normalised. To get the probability of tag assignment, these scores need to be 
  normalised by summation over all possible tags as shown in Table~\ref{tab:CRF-architecture}~(b). 
  Though the features are extracted locally using the function $f$, 
  the global normalisation captures long distance relationships in the word sequences.

  \begin{table}[ht!]
  \centering
  \footnotesize
  \begin{tabular}{ll}
   \multicolumn{2}{c}{\includegraphics[width=0.85\textwidth]{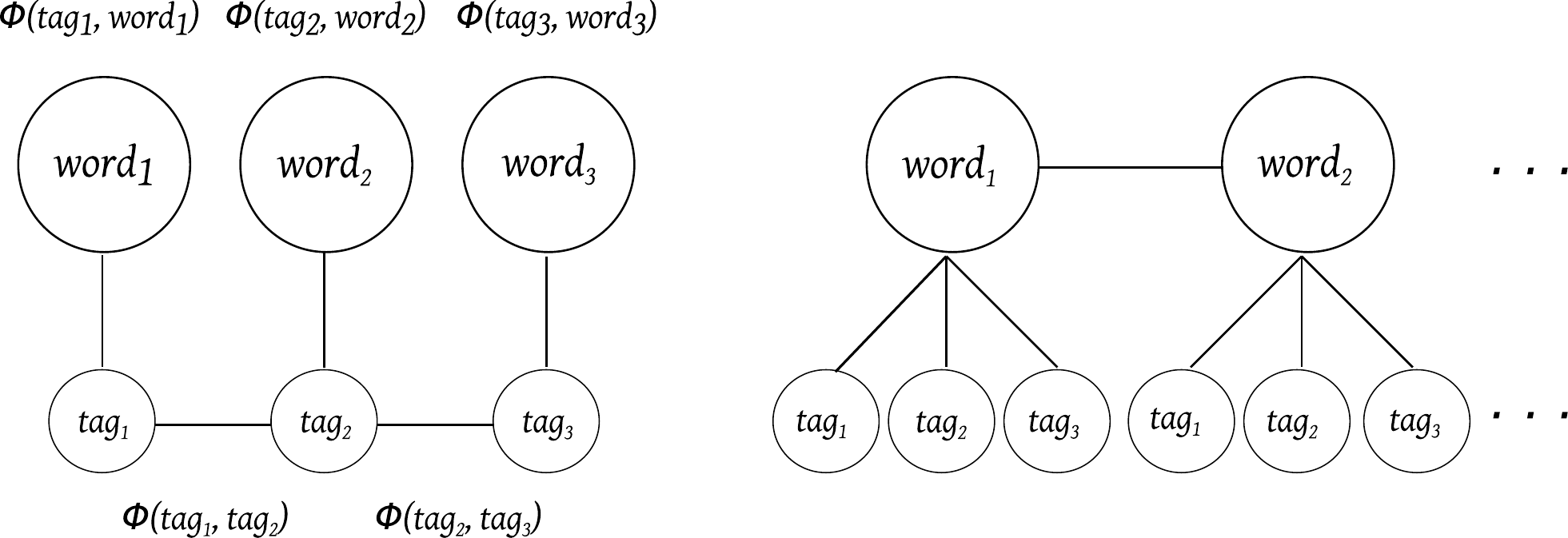}}\\
  (a)&(b)\\
  $P(tags|words) = \frac{1}{Z(words)}P(tags,words)$&$F=f(i,[null, tag_1, ..., tag_M]), \forall i \in [1,2, ...,N]$\\
  $\hat{P}(tags,words)=$&$\beta[M]=-LinearRegression(F)$\\
  $\prod_{j=1}^{M-1}\Phi(tag_j,tag_{j+1})\prod_{j=1}^{M}\Phi(tag_j,word_j)$
  &$\hat{P}(tags,words)=\Phi(tag_1, ...,tag_N|\x_1,...,\x_i, ....,\x_N) $\\
  $Z(words)=\sum_{tags}\hat{P}(tags,words)$&
  $P(tags|\x_1,...,\x_N)=\frac{\hat{P}(tags,words)}{Z(\x_1,...,\x_N)}$\\
  $\argmax_{tags \in \T}P(tags|words(Tweet_i))=$&
  $Z(\x_1,...,\x_N)= \sum_{tag_1=1:M}...$\\
  $\argmax_{tags \in \T}P(tags|\x_i)$&$\sum_{tag_N=1:M}\Phi(tag_1, ...,tag_N|\x_1,...,\x_i, ....,\x_N)$\\
  \end{tabular}
  \caption{Formalisation of sequence labelling task (a) a generic Conditional 
  Random Field (CRF), (b) LingPipe CRF implementation which we used in our pipeline.}
  \label{tab:CRF-architecture}
  \end{table}

  \paragraph{Training the CRF Model} The objective is to spot event and location terms in tweets.
  Identifying locations in a tweet is challenging as location references in the text are hard to recognise
  especially in the presence of non-standard abbreviations, spellings, and 
  capitalisation convention. 
  To address these challenges, we train the sequence model with the knowledge
  of locations from Open Street Maps (OSM)~\cite{Haklay:2008:OUS:1477057.1477249}.

  On the other hand, identifying event terms is even more challenging especially given the open
  domain nature of city related events. 
  To address this issue, background knowledge consisting of 
  domain dictionaries 
  are obtained from event reports of different web pages (see Table~\ref{tab:event-classes}),  \eg sport, weather and locations are such categories of events. The CRF is trained on short reports of such categorical event reports and then applied to our data for event terms name entity recognition. The result of this step (shown in Fig.~\ref{fig:NLP_pipeline}),
  forms the semantic embedding view and will be denoted with $\mathbf{\phi(\x)}$. This embedding can also be considered as a naive projection (embedding) of the output label space, $\calY$.

\subsubsection{\bf{CNN Word Tagging}}
\label{sec:CNN Event Tagging}
\begin{figure}[ht!]
  \centering
  \includegraphics[width=0.7\textwidth]{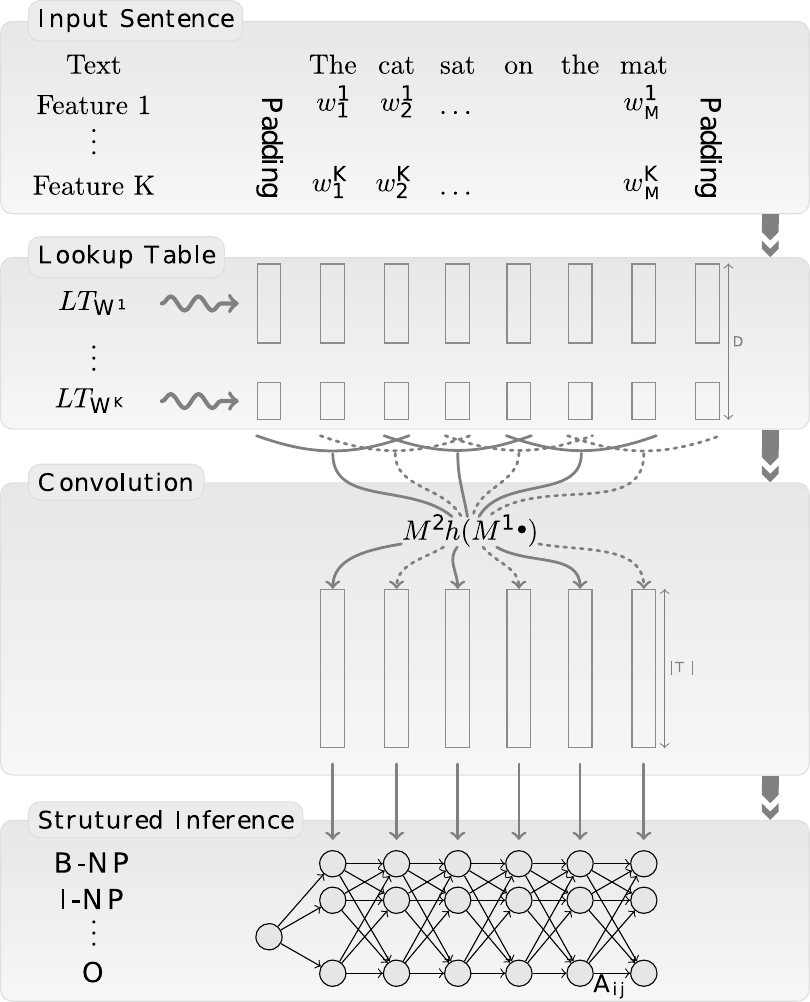}
  \caption{Convolutional Neural Network architecture, source: \cite{Collobert:2011:NLP:1953048.2078186}}
  \label{fig:CNN-view-architecture}
  \end{figure}

  The CNN model takes the input sentence and learns several layers of feature
  extraction that process the input tweets. The features computed by the deep layers 
  of the network are automatically trained by back-propagation.
  Fig.~\ref{fig:CNN-view-architecture} depicts the CNN network 
  architecture.

  \paragraph{Word-Level Feature Extraction} The CNN Word Tagging unit considers a fixed-sized word 
  dictionary~\footnote{Unknown words are mapped to a special unknown
  word. Numbers references are also mapped to a ``number'' word.} $\W$. Given
  a sentence of $M$ words ${w_1 , w_2 , . . ., w_M}$, where
  $w_m \in \W$, it is first embedded into a $D$-dimensional vector
  space where the index is taken from a finite dictionary of size $|\W|$, 
  by applying a look-up table operation:

  \begin{equation}
  LT_W (w_m) = W\Big( 0, \dots, 0, 1_{\text{at index $m$}}, 0, \dots, 0 \Big)^{T_2} = W_{w_m}
  \label{eq:CNN1}
  \end{equation}
  
  \noindent
  Matrix $W \in \mathbb{R}^{D \times|\W|}$ represents the parameters to 
  be trained in this look-up layer. 
  Each column $W_n \in \mathbb{R}^D$ corresponds to the embedding of the 
  $m^{th}$ word in the dictionary $\W$.

  Having in mind the matrix-vector notation in Eq.~\ref{eq:CNN1}, the look-up table 
  applied over the sentence can be seen as
  an efficient implementation of a convolution with a kernel width of size one. 
  Parameters $W$ are thus initialised
  randomly and trained as any other neural network layer. 
  These representations have been trained on the English Wikipedia corpus~\footnote{Available for download at http://download.wikipedia.org} affter using the Penn Treebank tokenizer~\footnote{Available at http://www.cis.upenn.edu/~treebank/tokenization.html.} and after removing all pragraphs containing non-roman characters and all MediaWiki markups.
  The extracted features contain syntactic and semantic information which appears to be useful for inference. 

  In practice, it is common that one wants to represent a word with more than one feature. 
  In such a scenario, the low-caps words and the "caps" feature: $w_m = (w_m^{lowcaps}, w_m^{caps})$
  can be used and to obtain this, one needs to apply different look-up tables for each 
  discrete feature ($LT_{W^{lowcaps}}$ and $LT_{W^{caps}}$), 
  and the final word embedding is formed by concatenating the output 
  of all these look-up tables:
  \begin{equation}
  LT_{W^{words}} (w_m) = \big(LT_{W^{lowcaps}}(w_m^{lowcaps})^{T_2}, LT_{W^{caps}}(w_m^{caps})^{T_2} \big)
  \label{eq:CNN2}
  \end{equation}
  For simplicity, we followed~\cite{Collobert:2011:NLP:1953048.2078186} suggestion  
  and considered only one look-up table.

  \paragraph{Sentence-Level Representation} Scores for all tags $T_2$ and all 
  words in the sentence are
  produced by applying a classical Convolutional Neural Network over the 
  look-up table embeddings obtained from Eq.~\ref{eq:CNN1}. More
  precisely, all successive windows of text (of size $K$) are considered by 
  sliding over the sentence, from position
  $1$ to $M$. At position $m$, the neural network of the structural inference step is trained with the vector $\x'_m$ resulting from the concatenation of the embeddings:

  \begin{equation}
  \x'_m = \big( W^{T_2}_{w_{m-(K-1)/2}}, \dots,  W^{T_2}_{w_{m+(K-1)/2}} \big)^{T_2}
  \label{eq:CNN3}
  \end{equation}

  The words with index exceeding the sentence boundaries 
  $(m − (K − 1)/2 < 1 \text{ or } m + (K − 1)/2 > M )$
  are mapped to a special padding word. As any classical neural 
  network, Collobert proposed architecture performs 
  several matrix-vector operations on its inputs interleaved with 
  some non-linear transfer function $f_2$. 
  It outputs a vector of size $|\T_2|$ for each word at position $m$, 
  interpreted as a score for each tag 
  in $\T_2$ and each word $w_m$ in the sentence:
  \begin{equation}
  s(\x'_m ) = \mathbf{M} ^2 f_2(\mathbf{M}^1 \x'_m ) 
  \label{eq:CNN4}
  \end{equation}
  where $H$ denotes the number of the hidden units and the 
  matrices $\mathbf{M}^1 \in \mathbb{R}^{H \times(KD)}$ and 
  $\mathbf{M}^2 \in \mathbb{R}^{|\T_2| \times H}$ 
  are the parameters to be trained on the network. 
  The “hard” version of the hyperbolic tangent function is utilised as the transfer function:
  \begin{equation}
      f_2(u) = \begin{cases}
		-1 & \text{if $ u < -1$} \\
		  u & \text{if $ -1 \leqq u \leqq 1$} \\
		  1 & \text{if $ u > 1$} \\
		\end{cases}
		\label{}
  \end{equation}

  Fine details of the adopted CNN architecture are explained in~\cite{Collobert:2011:NLP:1953048.2078186}.

\subsubsection{\bf{Multi-view Learning for tweet Annotation}}
\label{sec:Multi-view Learning}

  The dictionary-based NER approaches explained in previous sections are beneficial when a 
  text (\ie tweet) contains some event terms that is previously seen by the model in the predefined general English words.

  Given a sentence, these methods extract event terms by searching for 
  word sequences that match the 
  lexical entries, and create a token graph according to the word order.
  The next step is to estimate the score of every path using the weights of node and 
  edges estimated by training CRF (or CNN) and selecting the best path in 
  a joint learning model.

  While combining the two proposed NER tagging approaches can lead in performance enhancement, 
  when term ambiguity and variability are very high, specifically in the case of 
  tweets of short-sentence nature, dictionary-based Named Entity Recognition (NER)
  may not be an ideal solution even though large-scale terminological resources are 
  available~\cite{Sasaki:2008:MMN:1572306.1572318}.

  A common solution to enhance the performance would be the 
  addition of named entities to a Named Entity dictionary.
  However, in the case of multi-class annotation this might increase the risk of class confusions.
  Moreover, retraining of NER models is required to guarantee 
  achieving task specific class labels. 
  
  Consensus principle of multi-view learning as a joint learning model aims to maximise the 
  agreement on multiple distinct views. 
  Suppose the available Twitter data sample $\X$ has two views: the semantic view, $\phi(\x)$, which is 
  obtained from CRF+CNN NER word tagging and the syntactic view, $\theta(\x)$, which is 
  derived from CNN PoStagging. 
  An example $(\x_i , \y_i )$ is therefore viewed as 
  $(\theta(\x_i) , \phi(\x_i) , \hat{y_i})$, where $\hat{y_i}$ is the final 
  label assigned to sample $\x_i$.

  While the PoS tagging output of the CNN model will shed a light on 
  the grammatical structure of the text (\ie tweet) and possibly facilitates the 
  global inference on tweet's meaning, its NER location and organisation named 
  entity recognition output can be utilised for boosting the \textit{Location} name entity recognition of CRF model.

 Since retraining the last supervised fully connected layer of a convolutional neural network for adapting the learning for a new task is a common practice in deep learning,  we adopted the Restricted Boltzman Machine 
  (RBM)~\cite{conf/ciarp/FischerI12} formulation to perform the multi-view learning with the aim of event classification.
  
  The RBM is a Markov Random Field associated with a bipartite undirected graph. In the Bernoulli RBM, our focus
in this work, the visible and hidden variables are assumed to take values $(v, h) \in [0, 1]$. Each value encodes the probability that the specific feature would be active. Perceiving the RBM as an energy model~\cite{Adelson85spatiotemporalenergy,Fleet85EnergyModel}, the RBM feature learning encodes an input
  vector $\mathbf{v}$, using a vector of latent variables $\h=\sigma(W^T\mathbf{v})$. 
  Therefore each column
  of the weight matrix $W$ can be viewed as a filter which 
  corresponds to network's hidden variable $h_n$ in which $\sigma$
  is a non-linearity, such as the sigmoid, $\sigma(u)=\frac{1}{1+exp(u)}$.
   The weight parameters $W$ are then estimated through
   maximising the likelihood of the observations via Gibbs 
   sampling~\cite{Hinton:2006:FLA:1161603.1161605} 
   based on a set of training examples and the activity of the hidden unit. 
   The model is defined as the sum over the filter responses:
   \begin{equation}
   \bf{h}= W^T (F^T \bf{v}) \odot (F^T \bf{v})
   \label{eq:RBM-activity}
  \end{equation}
  where $\odot$ is element-wise product and the columns of $\bf{B}$ contain subspace projection filters that are learned along with $\bf{W}$ from data.

  \begin{figure}[ht!]
  \centering
  \includegraphics[width=0.7\textwidth]{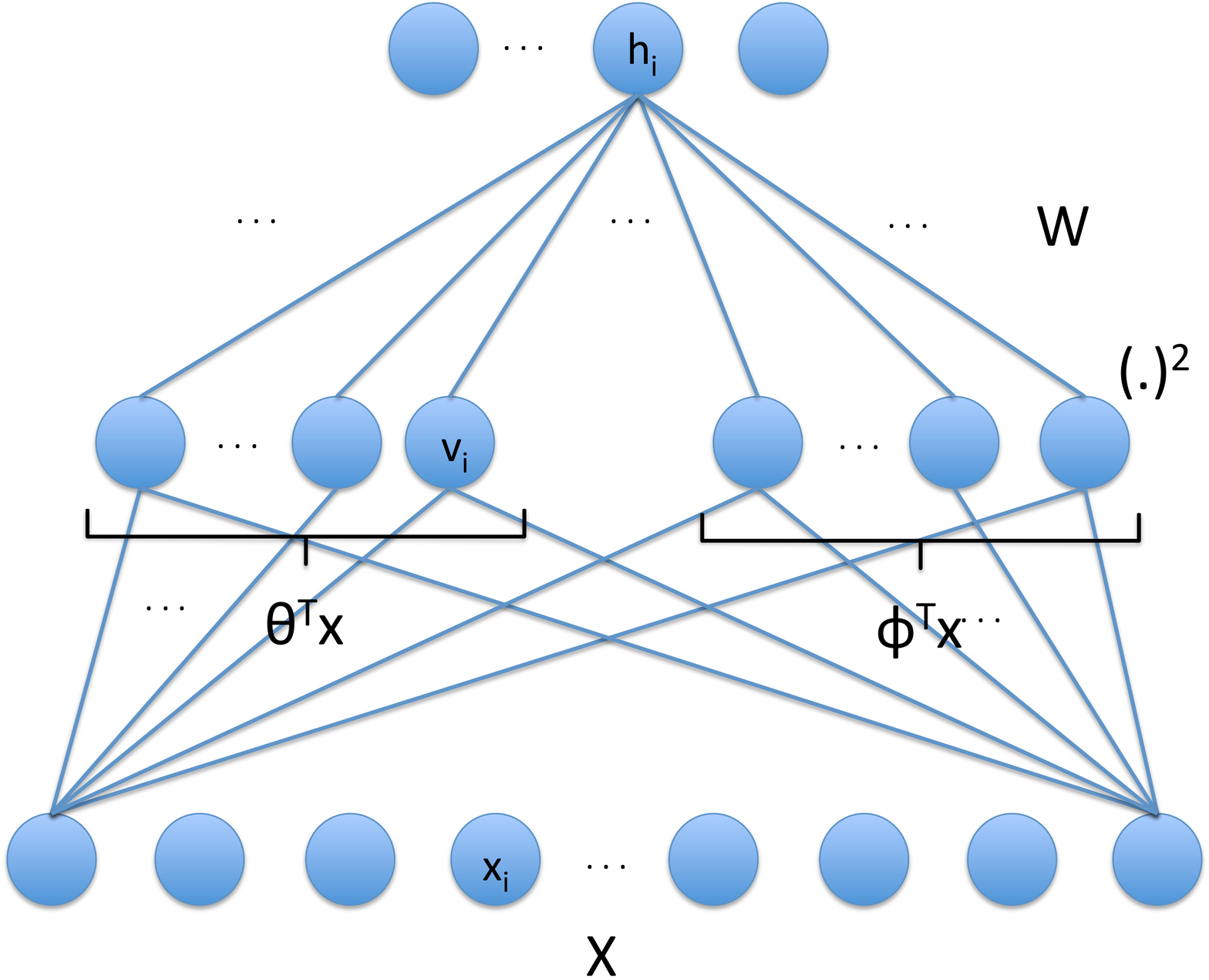}
  \caption{Modelling the multi-view learning through an RBM energy model applied to the concatenation of the semantic and syntactic embeddings of a sentence.}
  \label{fig:RBM-graph}
  \end{figure}
  
  When an energy model is applied to a concatenation of two views of a data, 
  a response that is closely related to the response of a multi-view sparse coding 
  model is obtainable. Inspired by Memisevic \cite{DBLP:journals/corr/abs-1206-4609}
   multi-view image correlation model, we can formulate our view fusion problem  
  via defining a compatibility energy function $E : \calX \times \calX' \rightarrow \mathbb{R}$
  that encode the relationship between the two views as shown in Fig.~\ref{fig:RBM-graph}. 
  We have modelled the visible layer of the RBM as the concatenation of two embeddings $\bf{B}^Tv = \Theta^Tx+ \Phi^Tx$ where $\Theta$ denotes the part of the filter $\bf{B}$ in Eq.~\ref{eq:RBM-activity} that is applied to input sentence $\x$ to extract its syntactic embedding and $\Phi$ denote the part of the filter $\bf{B}$ in Eq.~\ref{eq:RBM-activity} that extracts the semantic embedding of the sentence which can be perceived as a projection of the label space Y. $\bf{W}$ in this formulation is a $d \times |C|$ matrix representing the parameter space with $d=d_1 +d_2$ where $d_1$ and $d_2$ are the two view embedding dimensionality and $|.|$ represents the cardinality of the label set.
  
  Substituting $\bf{B}$ in Eq.~\ref{eq:RBM-activity} with a concatenation of view 
  projection matrices $\Phi$ and $\Theta$ 
  and $\mathbf{v}$ with $(\x,\y)_{d_1+d_2}$, the hidden unit activities 
  in the multi-view feature learning scenario, take the form:
  \begin{equation}
    h_i=\sum_j^d W_{ij} (\Theta^T\x+\Phi^T\x)^2= 2 \sum_j^d (\Theta^T\x)(\Phi^T\x) + 
    \sum_j^d W_{ij} (\Theta^T\x)^2+\sum_j^d W_{ij} (\Phi^T\x)^2
   \label{eq:multi-view-activity}
  \end{equation}
  
  In this formulation, the quadratic terms in Eq.~\ref{eq:multi-view-activity} are view-specific optimisation problems which have already been solved through the prior CNN and CRF training steps. 
 
  Having an estimate of the subspace projection matrices $\Theta$ and $\Phi$, we now just need to learn the weight matrix $\bf{W}$ from in-domain data (tweet instances). Given the fact that the activity of the second layer in the proposed architecture (see Fig.~\ref{fig:RBM-graph}) will be the concatenation of the projected views, the last layer can be trained by simply maximising the log-likelihoods over the training set.
Given a sentence $\x$, the network with energy function $E(v,h)=\sum_i\sum_j w_{ij}v_ih_j + \sum_i b_iv_i + \sum_jk_jh_j$ and parameter set $\xi=(\bf{W, b, k})$ computes a score $s^c_\xi(\x)$ for each event label $c \in C$.  In order to transform these scores into a conditional probability distribution of labels given the sentence and the set of network parameters, one can apply a softmax operation on the scores:
\begin{equation}
p(c|\x,\xi) = \frac{e^{s^c_\xi(\x)}}{\sum_{\forall i \in C} s^i_\xi(\x)}
\label{eq:softmax}
\end{equation}
Taking the log from two sides of the Eq.(\ref{eq:softmax}):
\begin{equation}
\log p(c|\x,\xi) = s^c_\xi(\x) - \log (\sum_{\forall i \in C} s^i_\xi(\x))
\label{eq:log-softmax}
\end{equation}
One can then use the gradient decent theorem to minimise the negative log probability with respect to $\Theta$. 
The back-propagation algorithm is a natural choice to efficiently compute gradients of the network architecture as stated in~\cite{Collobert:2011:NLP:1953048.2078186}.
  
  \subsection{Tweet Impact Estimation}
  \label{sec:Tweet Impact Calculation}
  Many factors are contributing in reliability of an event extracted from Twitter data;
  there might be multiple references to the same event and the event's extracted location might
  be different from the location where the tweet is published.
  
  To capture this, we define the tweet impact factor 
  as the product of event severity 
  and event likelihood scores following~\cite{StatisticalMethods4Immunogenicity2015}:
  \begin{equation}
   \hat{e_{impact}}=score_{s} \times score_{l}
   \label{eq:impact-cal}
  \end{equation}
  \noindent
  where the event severity score, $score_{s}$, is calculated following the spatio-temporal
  event grouping approach of~\cite{Pramod-ACM-2015} referred as Thematic Coherence. The
  Thematic Coherence approach considers events with similar entities, reported within a grid $g_i \in G$ (where
  $G$ is a set of all grids in a city) and time $\delta t$ as multiple references of 
  the same event and reports the severity score as the total number of events falling in this criteria.
  
  In our evaluations, we have fixed the time $\delta t$ to five minutes and unlike 
  \cite{Pramod-ACM-2015} who used the tweet's geo-tag for computing the thematic coherence,
  upon existence we have utilised the extracted event location (the output of the multi-view tweet 
  annotation described in section~\ref{sec:Multi-view Learning}) 
  along with the predicted event type for grouping and computing the event
  severity scores.
  
  We have also formulated the event likelihood score computation as follows:
  \begin{equation}
   score_{l}= 1-\frac{Dist(location,g_{C})}{Dist(g_{BB(0)},g_{BB(1)})}
   \label{eq:likelihood-score}
  \end{equation}
  \noindent
  where the $Dist(g_1,g_2)$ function measures the Vincenty distance~\cite{Journal/geo/Karney13} between 
  two geolocation coordinates. The $(latitude, longitude)$ pairs, 
  $g_{CC}$, $g_{BB(0)}$ and $g_{BB(1)}$ are corresponding  to city-specific centroid and bounding box 
  information, respectively which are estimated using the Flickr's Geo API 
  Explorer~\footnote{https://www.flickr.com/places/info/44418}. 
  Therefore, these values will be set as 
  $g_{C}=(-0.1280,51.5077), g_{BB(0)}=(-0.5103,51.2868) , g_{BB(1)}=(0.3340,51.6923)$ 
  for the London city.
  The event likelihood score in practice will assign more impact to the events which are reported closer to the city centre.

\subsection{Similarity Analysis Graph Representation}
\label{sec:CorrGraph}
For the similarity analysis, we narrowed down our focus to the event classes which enabled the access to the authority event records, namely the Transport and traffic reports and scheduled sociocultural records. To do this, we collected officially registered traffic reports from London open data store portal and parsed the Time Out London webpage to get a list of scheduled sociocultural events taking place in London along with their timestamp and locations.

\begin{figure}[ht!]
\centering
\includegraphics[width=.7\columnwidth]{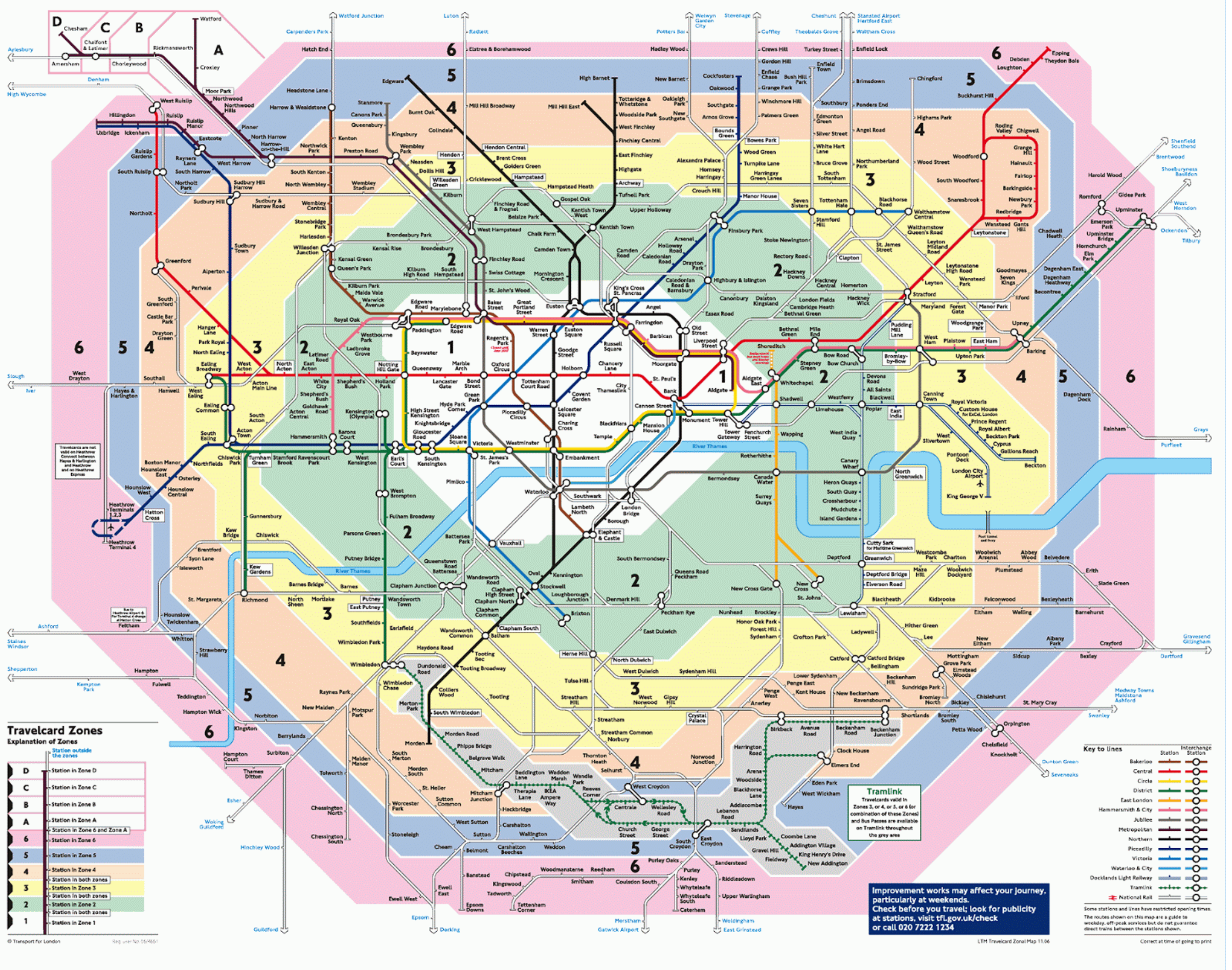}
\caption{Colour-coded map of the London zones showing the spectral topography of the city. Note that the use of colours on the map is only to depict the spectral distribution of the city locations.}
\label{fig:London_Topography}
\end{figure}

Two graph structure have been considered to represent the spatial distribution of the traffic and sociocultural records. Let assume that, $G^T=\{n^T_1,n^T_2, ...,n^T_{|T|}\}$ and $G^{SC}=\{n^{SC}_1,n^{SC}_2, ...,n^{SC}_{|SC|}\}$ representing the traffic and sociocultural record graphs respectively where $|.|$ denotes the cardinality of each record set. The edge values in these graphs are associated with the pair-wise Euclidean distances between the nodes which are in 3D coordinates and are denoted with feature vectors $\bf{n_i}=(x_i,y_i,z_i,t_i,e_i)$. The first three variables, $x,y,z$, represent the spatial coordinate of a point after polar to Cartesian conversion and $t_i, e_i$ are the event timestamp and event type, respectively. We employed these two graphs for detecting the nearest node to each of the automatically annotated Twitter events. Moreover, we have taken into account the spatio-spectral topography of London city presented in Fig.~\ref{fig:London_Topography}.
As one can note, the spectral structure of the city infers a spectral weighting of computed distances. Meaning that, the farther we move away from the city centre, the distance turns to be inversely prominent. Taking this into account, given that seven target graphs each represents the twitter events of each class and simplifying the location vector of each node with $P_i=(x_i,y_i,z_i)$, we then formulated the graph dissimilarities with respect to authority graphs (\ie traffic) as follows:
\begin{equation}
\hat{D_c} = \Sigma_{i=1:|c|} \min(d(P^{G^c}_i,P^{G^T}_j)/\lambda_j)
\label{eq:dissimilarity}
\end{equation} 
where $d(.)$ represents the Euclidean distance between two points. The points superscripts, $G^c, G^T$, shows the graph memberships, and $|c|$ denotes the cardinality of tweets, which are classified as event type $c$ and with $P_{centre}$ being the Cartesian conversion of the city centre geo-coordinates~\footnote{Following flicker, (-0.127, 51.507) is considered as the (longitude, latitude) pair describing the city centre}. The parameter $\lambda$ is formulated as  $\lambda_j=d(P^{T}_j,P_{centre})$. In practice, the value of the parameter imposes higher weight to close-to-centre events compared to off centre events.

\section{Experimental Setup and Results}
\label{sec:exp}

Our experimental objective is to evaluate the proposed framework performance and its extendability for tweet classification where the data is collected from new locations and at varying time with respect to training data. To showcase this, 
we conducted experiments on textual Twitter data collected from two
geographically different locations: San Francisco Bay area and London.

Our objective in the following evaluations are three-fold: i) to quantify the extent to which our framework can extract city events from Twitter where we compare our approach with the state-of-the-art baselines~\cite{Pramod-ACM-2015,Ritter:2012:ODE:2339530.2339704} on \textit{San Francisco} data, ii) to evaluate the performance boost of the proposed MV-RBM approach for sentence inference rather than just word tagging by testing the model on locally collected dataset from London; iii) finally to perform  similarity analysis and study how well the Twitter extracted events are matching with authority reports.

\subsection{Datasets}
\label{sec:datasets}
To make the evaluation, we constrain our experiments to the domain of city
related events. The proposed approach is generic enough to
be applied to any other cities for which the Twitter data is available. The final aim in the proposed pipeline is to assign one (or multiple) label(s) to each tweet out of a set of city event classes 
\{Crime, Transportation (Trans.), Cultural event, Sport, Social event, Food, Weather and Location\}.
We leverage the open domain knowledge available for a city, specifically, vocabulary
related to each of these categories from official and authorised web reports 
as summarised in Table\ref{tab:event-classes}, 
\ie Transportation (Trans.) vocabulary is constructed using phrases that are taken from 
http://511.org~\footnote{\url{http://511.org}} web page,  
the Open Street Map (OSM)~\footnote{\url{http://www.openstreetmap.org}} of the cities is used for extracting the city location terms and 
the Wikipedia cultural activities hierarchy~\footnote{\url{http://en.wikipedia.org/wiki/Category:Cultural_events}} 
is utilised for constructing the Cultural event terms. 

\subsubsection{Data Collection through Twitter Streaming API}
\label{sec:data-collection}
The Twitter data which is used in this study has been collected 
via Twitter Streaming API which allows 
searching for keywords, hash tags, user Ids and geographic bounding boxes simultaneously.
The \textit{filter} API facilitates the search by providing a continues stream 
of tweets matching the search criteria. Three key parameters are used for the search:
\begin{itemize}
\item Follow: a comma-separated list of user Ids to follow, which 
returns all of publicly published tweets in the stream.
\item Track: a comma-separated list of keywords to track. 
\item Location: a comma-separated list of geographic bounding boxes 
containing the coordinates of the southwest point and the northeast point 
as a $(longitude, latitude)$ pair.
\end{itemize}

Twitter Streaming API limits the number of parameters which can be supplied in one request. 
Up to $400$ keywords, $25$ geographic bounding boxes and $5000$ user Ids 
can be provided in one request. In addition, the API returns all matching 
tweets up to a volume equal to the streaming cap where the cap is currently
set to $1\%$ of the total current volume of tweets published on 
Twitter~\cite{TwitterDataAnalytics2013}.

We used the \textit{San Francisco} Twitter data collected by~\cite{Pramod-ACM-2015} 
for a period of four months (Aug 2013 to Nov 2013). 
While the original dataset contained over 8 million tweets for this time period,
the authors sub-sampled the data, resulting in a test dataset of size 500 tweets for testing their trained model. 
We have used the same test dataset for our comparative evaluations. This dataset is referred to as \textit{San Francisco}.
 Additionally, we have collected data from London using all API parameters (Location bounding box, tracking and following official news agency user names and user Ids) at two different timestamps, referenced in the remaining of this paper as $London_1$ and $London_2$. 
 The $London_1$ data is composed of 3000 tweets collected between $15^{th}$ and $31^{th}$ of May 2015 and manually cleaned and annotated for training and testing the MV-RBM model. The manual annotation results undergo a second investigation for ensuring their consistency and validity. We have asked a group of technical users, who work in the field of smart cities to peer-review the validation of the annotations.
The $London_2$ data is collected on $3^{ed}$ of February 2016 and is of size $1.1MB$. In section~\ref{sec:eval2}, we used this dataset to examine the Twitter extracted event similarity with the road sensor data and the scheduled events that are parsed from the Web.
 
\begin{figure}[t!]
\centering     
\subfigure[]{\label{fig:temporal-tweet-dist}\includegraphics[width=0.5\textwidth]{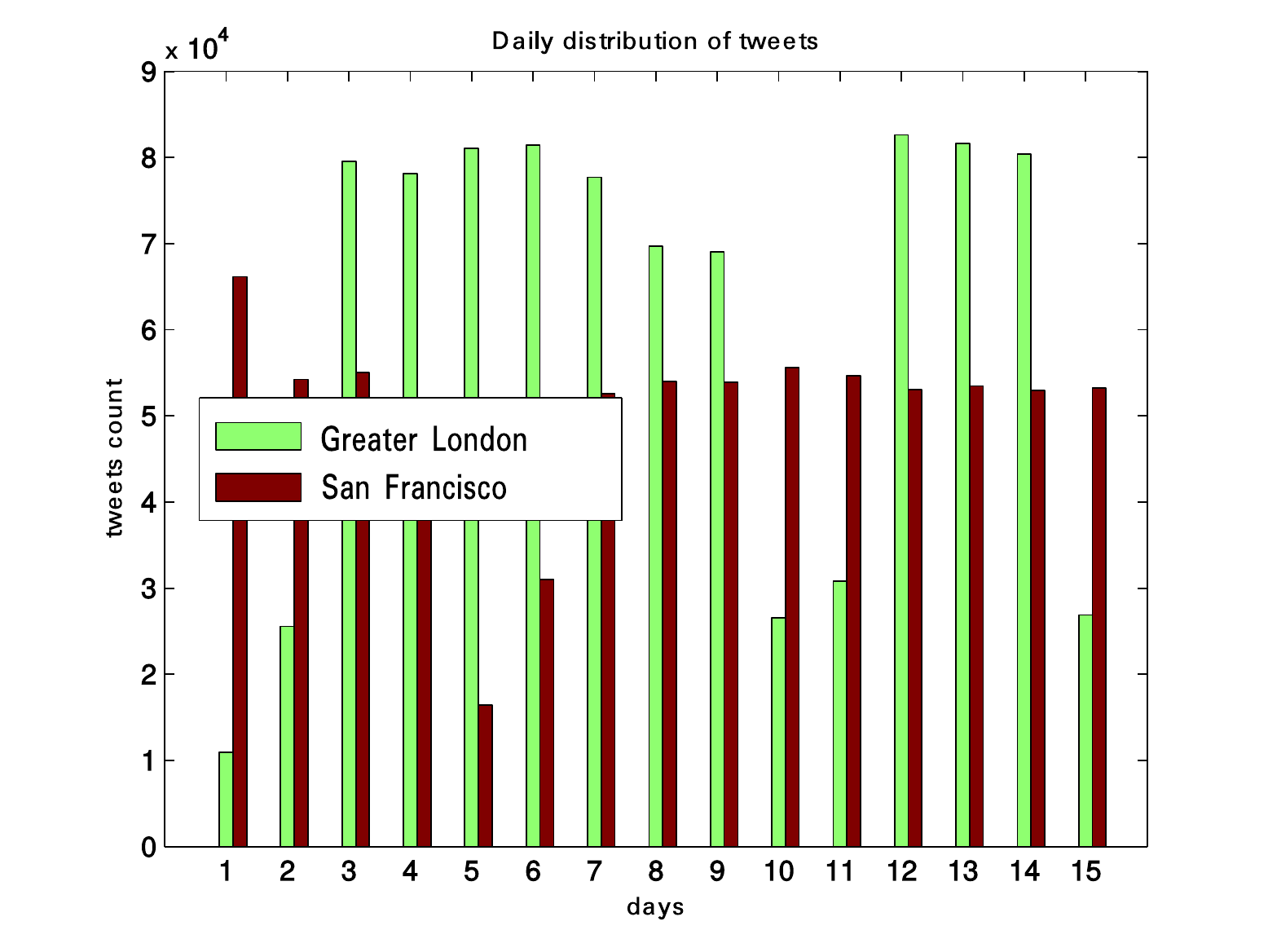}}
\hfil\raisebox{.08\textwidth}{
\subfigure[]{\label{fig:AnnotationTool}
\includegraphics[width=0.45\textwidth]{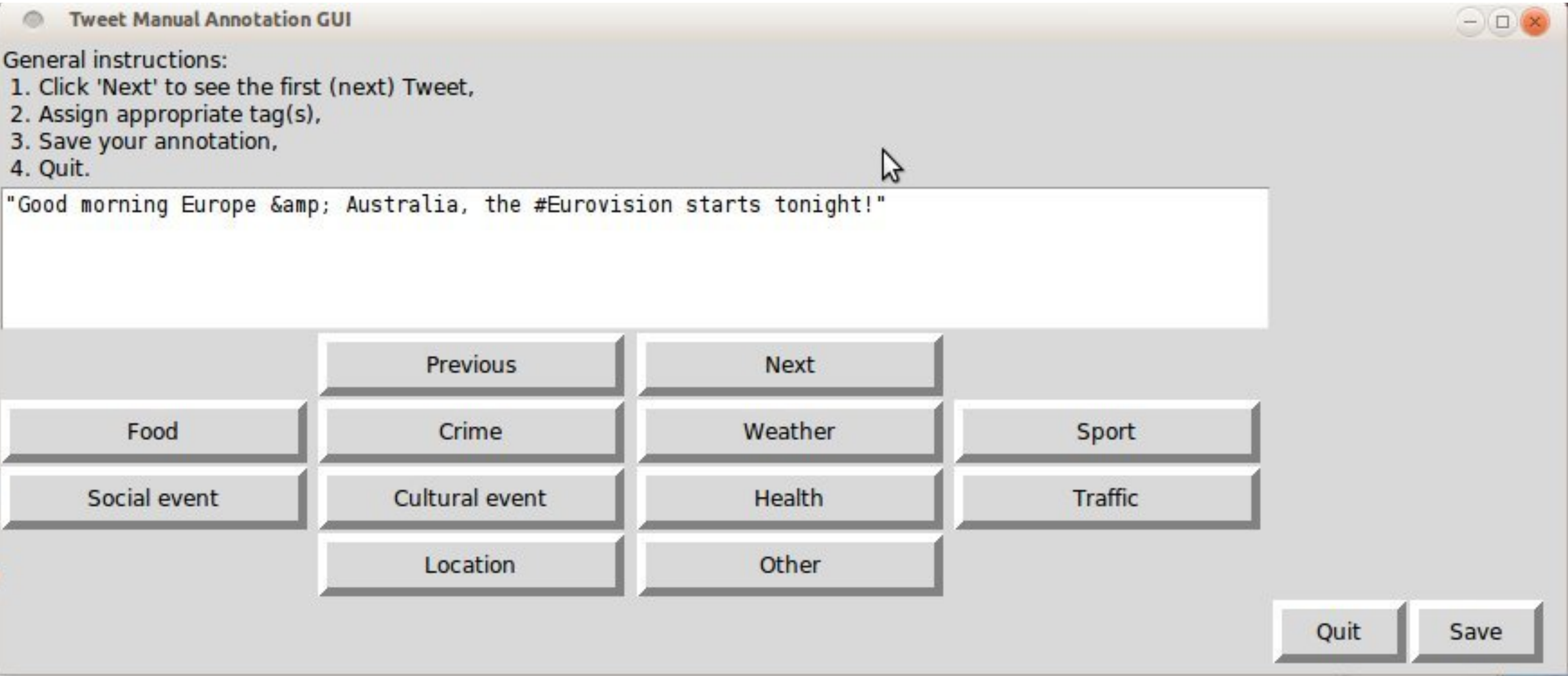}}}
\hfil
\caption{a: A comparison of temporal distributions of daily tweets in two different cities within 15-days time period: $London_1$ for the time period of 15/05/2015 to 01/06/2015 and \textit{San Francisco} for the time period of 01/08/2013 to 15/08/2013., b: View of the Tweet annotation tool.}
\end{figure}

Temporal distribution of daily tweets collected from \textit{San Francisco} and $London_1$ datasets are shown in Fig.~\ref{fig:temporal-tweet-dist}. 
\subsubsection{Ground-truth Annotation Tool}
\label{sec:Tweet-GT-annotation}
To facilitate the ground-truth annotation of $London$ data, we have developed a GUI tool.
A view of this tool is represented in Fig.~\ref{fig:AnnotationTool}~\footnote{The annotation tool
is available for downloading at: \url{http://goo.gl/UBTKQp}}. 

For this study, we have used this tool to annotate the 3000 $London_1$ tweets for constructing a training dataset of size 2000 and a test set of size 1000 tweets for training and evaluating our model.
We have asked a group of seven technical users who work on smart city research in our team to peer-review the annotations.
Following criteria have been asked to be considered during the annotations:
\begin{itemize}
 \item The user is asked to assign tweets to one or more classes of events, having in mind the potential effect
 of the event on city daily pattern
 \item The \textit{``Social''} class is the class which includes voting, election, 
 protest and other city related group activities
 \item The class \textit{``Other''} is an indicator for tweets which either their contents can not be 
 associated with any of the provided classes (\ie personal messages and opinion sharing) or
 they are hard to be understood (\ie ambiguous notes and texts that are hard to 
 follow in the absence of background knowledge).
\end{itemize}
The result of the ground-truth annotation is used for training the proposed multi-view learning model and to 
 perform evaluations.

\subsection{Performance Evaluation I: Word Level Annotation}
\label{sec:eval}
In this section we will evaluate the three proposed city-event classification algorithms performance.
We present detailed performance measures on the two datasets collocated from two different cities and 
show how different steps of the proposed multi-view learning model can 
help to achieve an enhanced performance considering all different views of the same data.
 
 \subsubsection{CRF Tweets Tagging Evaluation}
 \label{sec:CRF Tweets Tagging Evaluation}
 In order to evaluate our CRF name entity recognition model and 
 to assess the efficiency of the proposed class-conditioned report catalogues, 
 we compared the performance of our model against  two state-of-the-art approaches: $B_1$~\cite{Ritter:2012:ODE:2339530.2339704} and $B_2$~\cite{Pramod-ACM-2015}. Unlike our supervised event extraction, the two baseline approaches are unsupervised. Therefore, we had to combine all event classes of our framework against the non-event (other) class and mainly focus on the coverage of events and the event locations.
 We have used the $San Francisco$ dataset for this comparison as the baseline approaches had been evaluated using this dataset.
 Table~\ref{tab:CRF-SanFrancisco-PramodvsOurs} shows this evaluation results. 
\begin{table}[ht]
\caption{Comparisons of our CRF dictionary tagging vs. the baseline 
($B_1$ and $B_2$) methods of Anantharam \etal
and our universal English CRF dictionary tagging approach.}
\begin{tabular}{|c|c|c|c|c|c|c|c|c|c|c|c|c|}
\hline
    \multirow{2}{*}{}&\multicolumn{3}{c|}{Other}&\multicolumn{3}{c|}{Location}&\multicolumn{3}{c|}{Events}&\multicolumn{3}{c|}{Precision}\\ \hline
 &$B_1$&$B_2$&Ours&$B_1$&$B_2$&Ours&$B_1$&$B_2$&Ours&$B_1$&$B_2$&Ours\\ \hline
 Other&3936 &4267&4227&590 &175&68&178 & 9&40&0.84& 0.96&\bf{\blue{0.97}}\\ \hline
 Location&336 &76&46&459& 983&972&20 & 2&0&0.56 &0.93&\bf{\blue{0.95}} \\ \hline 
 Event& 26& 14&29&4 &0&13&70 &85&225&0.7 &\bf{\blue{0.86}}&0.84\\ \hline
 Recall&0.91&\bf{\blue{0.98}}&\bf{\blue{0.98}}&0.43&0.85&\bf{\blue{0.92}}&0.26& \bf{\blue{0.88}}&0.85&\cellcolor{gray}&\cellcolor{gray}&\cellcolor{gray} \\ \hline
\end{tabular}
\label{tab:CRF-SanFrancisco-PramodvsOurs}
\end{table}

We found that our CRF-NER model performed equally well as the best performing baseline model, $B_2$, recalling for $95\%$ vs $93\%$ precision on Location terms detection and $84\%$ vs $86\%$ precision on event detections.
Note that while $B_2$~\cite{Pramod-ACM-2015} method had been trained on a large corpus of approximately $8$ million tweets collected from \textit{San Francisco}, out CRF model was only trained on generic city-independent report catalogues of Table~\ref{tab:event-classes} which means we did not provide domain (city) specific prior knowledge for training our CRF model. Instead, we used the CRF-NER tagging approach that is more flexible due to benefiting from a more generic set of conditional class terms. This enabled us to remove any geographical or temporal bias. However, our model performed as well as the model, which is specifically trained for a controlled domain (San Francisco Bay area) with access to official traffic reports of a given time period. Overall, we developed a model that is more flexible and adaptable, while producing results that were as good as the baseline approach. Therefore, our approach can be used for other cities with potentially varying event distribution.

To demonstrate the granular performance on each of our defined city event classes, we have presented the NER tagging results in terms of confusion matrix in Table~\ref{tab:CRF-SanFrancisco}.
 In order to obtain the ground truth NER tags for this confusion matrix, 
 we have used the same tagging schema as of proposed in~\cite{Pramod-ACM-2015}, 
 with the B- and I- prefixes referring to beginning and intermediate 
 tags respectively where exist multiple consecutive tags in an entity phrase~\footnote{Note that in this way each detected event (location) phrase in a given tweet might be composed of multiple terms and thereby multiple tags.}.
 
  \begin{table}[ht]
\caption{Evaluation results of the CRF dictionary based annotation on \textit{San Francisco} data subset}
\scriptsize
\begin{center}
\begin{tabular}{|p{12mm}|c|p{5mm}|p{7mm}|p{4mm}|p{7mm}|c|p{5mm}|p{4mm}|p{7mm}|p{5mm}|c|}
    \hline
    \multirow{11}{*}{\begin{tabular}{c} \bf{\blue{CRF dic.}} \\ \bf{\blue{Tagging}}\\\end{tabular}}&\multicolumn{9}{c}{\bf{\blue{Ground-truth Labels}}}&&Total\\\cline{2-12}
    &\cellcolor{gray}&Crime&Cultural&Food&Location&\begin{tabular}{c}Other \\(non-event)\\\end{tabular}&Social&Sport&Weather&Trans.&\cellcolor{gray}\\\cline{2-12}
    &Crime&12&0&0&0&4&0&0&0&0&16\\\cline{2-12}
    &Cultural&0&39&0&0&1&0&0&0&0&40\\\cline{2-12}
    &Food&0&0&63&4&1&0&0&0&0&68\\\cline{2-12}
    &Location&0&0&0&974&46&0&0&0&0&1020\\\cline{2-12}
    &\begin{tabular}{c}Other \\(non-event)\\\end{tabular}&1&9&6&68&4227&5&4&2&6&4335\\\cline{2-12}
    &Social&0&1&0&4&17&27&0&0&0&49\\\cline{2-12}
    &Sport&0&0&0&1&5&0&19&0&0&25\\\cline{2-12}
    &Weather&0&0&0&2&0&0&0&21&0&23\\\cline{2-12}
    &Trans.&0&0&0&2&1&0&2&0&44&49\\\cline{2-12}
	\hline
    Total&\cellcolor{gray}&13&49&69&1055&4302&32&25&23&50&\cellcolor{gray}\\
    \hline
    Recall&\cellcolor{gray}&0.92&0.79&0.91&0.92&0.98&0.84&0.76&0.91&0.88&\cellcolor{gray}\\
    \hline
    Precision&\cellcolor{gray}&0.75&0.97&0.93&0.95&0.97&0.55&0.76&0.91&0.90&\cellcolor{gray}\\
    \hline
    F-measure&\cellcolor{gray}&0.83&0.88&0.92&0.94&0.97&0.67&0.76&0.91&0.89&\cellcolor{gray}\\
    \hline
    1vs.All Acc.&\cellcolor{gray}&0.8&0.79&0.75&0.96&0.95&0.77&0.78&0.94&0.83&\cellcolor{gray}\\
    \hline
\end{tabular}
\end{center}
\label{tab:CRF-SanFrancisco}
\end{table}
 
 In Table~\ref{tab:CRF-SanFrancisco} we have reported Precision, recall and f-measure scores for class dependent word tagging for the same dataset. 
 The one vs. all Tweet classification accuracies are also computed via 
 dividing the true positive rate by total number of samples of a class. 

The noteworthy is the slight difference between the one vs. all Tweet classification accuracy 
and word-tagging recall rate in Table~\ref{tab:CRF-SanFrancisco}. 
This in fact is caused by the intrinsic difference in word-level tagging vs. Tweet annotation made by human experts where whole Tweet meaning has taken into account and inference is involved. 
Investigating the Ground-truth Tweet annotations by expert users
depicts that annotation differences are occurring under two general circumstances.
The first source of such slight differences is where CRF-based label prediction mistakes are initiated from the assumptions made in sentence class label associations.
As also reported by Anantharam \etal~\cite{Pramod-ACM-2015}, subtle changes in context result in diverse interpretation of Tweet and subtle difference in location and event references and can cause loss of precision. 
An example is where tweets are assigned to class \textit{``Other''}. This class association is 
based on absence of any non-other class word tags within a Tweet. This means that if a 
word in a Tweet is tagged as \textit{``Location''} the Tweet will be labelled as 
\textit{``Location''} regardless of its global meaning.

The second, is where wrong Ground-truth labels are assigned to tweets due to experts' lack of common-knowledge. 
This itself is of two origins:  i) people normally are not aware of all events 
taking place and also of all locations existing in a city, 
 and ii) there are oddly phrased tweets which understanding them is quite 
challenging without following tweets on a specific topic which are tweeted by a specific user - the user Ids 
were not included in our data due to user privacy policy.).
While we have minimised the probability of such mistakes caused due to lack of individuals general knowledge with our peer-reviewed annotation scheme, the second cause remains intact as following historical data from a user is not allowed on Twitter stream API. This user's ground-truth annotation mistake, 
indeed demonstrates the necessity of an automated machine 
annotation model.

In order to partially tackle the failures caused by the CRF-dictionary annotation, we have proposed an alternative to Anantharam et al. CRF learning by boosting the CRF dictionary knowledge view through utilising a CNN generated view which jointly aims at enhancing the Location word tagging and providing words grammar roles. The two views of the data are then fed to a multi-view learning framework to enable a sentence-level reasoning and classification.
In the next two sections, we will further investigate and evaluate these claims.

 \subsubsection{CNN-enhanced LOCATION Tagging Evaluations}
 \label{sec:CNN Event Annotation Evaluations}
 As mentioned earlier, two alternative solutions can help 
 in enhancing the word level Tweet annotation: i) boosting the word tagging through fusion of multiple approaches and 
 ii) training a model which considers a sentence level reasoning for Tweet annotations (\ie classification) rather than solely relying on event-tag occurrences.
 
 To achieve the former enhancement, we used the CNN derived tags which boosts the 
 tagging accuracy of the \textit{LOCATION} class from $0.96$ which was previously reported in Table~\ref{tab:CRF-SanFrancisco} to $0.99$. 
 This \textit{Location} tagging enhancement is important in our framework since it 
 will enable us to assign more accurate locations to each extracted event rather than assigning the events to their tweeted locations~\footnote{Tweet's 
 Geo-tags in most of the occasions is different from actual event's location as people rarely publish 
 their thoughts about an event exactly in the actual venue of that event.}
 which was reported in previous studies~\cite{Pramod-ACM-2015}. 
 
 To better evaluate our claims we have also tested our proposed word tagging approaches 
 on a more realistic dataset collected from London referenced as $London_1$.
 Unlike the \textit{San Francisco} test data~\cite{Pramod-ACM-2015}, the $London_1$ data has not been cleaned prior to evaluations. Though basic pre-processing steps such as tokenizing and stop word removals have been included in the pipeline.
 The dataset is divided into two sub-corpora for training and testing
 the fused NER model. The results are reported in Table~\ref{tab:CNN-CRF-London}. 

\begin{table}[ht]
\caption{Evaluation results of the CRF-dictionary vs. CNN-enhanced tagging on $London_1$ 
city and \textit{San Francisco} datasets}
\scriptsize
\begin{center}
\begin{tabular}{|c|c|c|c|c|c|c|}
    \hline
    \multirow{2}{*}{\begin{tabular}{c}Location tagging\\performance\\\end{tabular}}&\multicolumn{3}{c|}{CRF-dictionary Tagging}&\multicolumn{3}{c|}{CNN-enhanced LOCATION Tagging}\\ \cline{2-7}
     &Recall&Precision & F-measure&Recall&Precision & F-measure\\ \hline
    \textit{San Francisco} data&0.96&0.93&0.94&\bf{\blue{0.99}}&0.87&0.93\\\hline
    $London_1$ data&0.49&0.83&0.61&\bf{\blue{1.00}}&0.43&0.59\\\hline
\end{tabular}
\end{center}
\label{tab:CNN-CRF-London}
\end{table}

Comparing this results with the performance measure on \textit{San Francisco} data
shows a slight degradation in Tweet event annotation performance for all classes. 
The main reason is that the \textit{San Francisco} data which had been used in previous experiments 
had gone through additional data cleanings (see details in~\cite{Pramod-ACM-2015}) prior to testing which 
in turn helped in leveraging the final performance.
Moreover, the effect of the CNN tagging which enhances the \textit{San Francisco} 
Location term tagging is slightly controversial in the case of $London_1$ data. 
While \textit{San Francisco} Twitter users~\footnote{One should also note that the San Fransicso bay areas is populated by industrial companies and 
organisation which in practice will lead in more harmonic text patterns of 
tweets published within its relative bounding box.} 
were more frequently and correctly used the ``@''
and ``\#'' characters for referring to locations and organisation names, 
the $London_1$ twitters have been observed to ignore these rules more frequently.

\subsection{Performance Evaluation II: Multi-View Learning}
\label{sec:eval2}
Benefiting from the name entity word tags and word syntactic roles assigned by CNN, in this section we will investigate how the two views can be trained simultaneously to realise an enhanced tweet class inference. As one might have noticed in Table~\ref{tab:CRF-SanFrancisco-PramodvsOurs}, the cardinality of tweets belonging to  \textit{``Other''}  category makes the Twitter corpora quite unbalanced for a supervised learning task. To tackle this issue, we have sub-sampled the subset of data of \textit{``Other''} class prior to our multi-view training step. This data sub-sampling step after the CRF NER tagging sounds plausible as the absence of any name entity tags from all non-other classes in a tweet can be assumed as a course classification of that tweet as class \textit{``Other''}. 
Table~\ref{tab:MV-Learning-London} 
demonstrates the performance evaluation of the proposed multi-view approach (presented in Sec.~\ref{sec:CNN Event Annotation Evaluations}). The performance is reported in terms of one vs. all class accuracies and the multi-view approach shows $5\%$ improvement in the performance compared to the single view model that classified tweeted events according to the CRF NER word tagging.

\begin{table}[ht]
\caption{Numerical results of multi-view learning evaluation on $London_1$ data}
\scriptsize
\begin{center}
\begin{tabular}{|c|c|c|c|c|c|c|c|}
    \hline
    1 vs. All Class Acc.& Crime& Cultural& Food& Social&Sport&Weather&Trans.\\ \hline
  \begin{tabular}{c}CRF $+$ CNN \\ word tagging\\annotation\\\end{tabular}& 0.53&0.36&0.35&0.16&0.52&0.8&0.67 \\\hline
  \begin{tabular}{c}MV-RBM\\ annotation\\\end{tabular}&\bf{\blue{0.6}}&\bf{\blue{0.37}}&\bf{\blue{0.48}}&\bf{\blue{0.21}}&\bf{\blue{0.54}}&0.8&\bf{\blue{0.69}}\\\hline
\end{tabular}
\end{center}
\label{tab:MV-Learning-London}
\end{table}

However, one can note a degradation of performance when it is 
compared with \textit{San Francisco} data classification results (last row of Table~\ref{tab:CRF-SanFrancisco}).
As mentioned before, this degradation was expected, as unlike the \textit{San Francisco} data 
(the collection and filtering described in~\cite{Pramod-ACM-2015}), 
non of the London datasets has been cleaned prior to the system annotation.

We obtained a lower performance on \textit{``Social'', ``Cultural''} and \textit{``Food''} topics compared to other topics. This can be explained by two reasons: class conditional dictionary term similarities in \textit{``Social''} and \textit{ ``Cultural''} categories and incomplete class conditional dictionaries in \textit{``Food''} category.

Although extra cautions have been taken for constructing  
class-specific dictionaries however, there were some terms and phrases which contributed in more than one dictionary. This has made the event extraction process more challenging in such scenarios. 

The effect of the dictionary problem can be reduced by increasing the training data from the categories that provide less accurate results. However, we did not tend to bias our data by providing 
more training samples of these categories and reported the results based on fair number of annotated tweets for our categories.
An example is in tweet \textit{``Rainbow food @ The Good Life Eatery''} 
where the \textit{rainbow food} phrase will be tagged 
as \textit{``B-Weather''} and \textit{``B-Food''} using the single-view CRF NER tagging while the actual tag should have been 
\textit{``B-Food''} and \textit{``I-Food''}. Such mistake will be resolved, once the 
NER tagging outcome is jointly weighted with the PoS tags derived from CNN, in our proposed multi-view learning step.

The multi-view training step can in fact provide more flexibility in 
expanding class-specific dictionaries. However, adding more terms will require the model to be trained with a 
larger size training data and will cause higher time and computational complexity and requires more manual 
annotation effort.

\subsection{Performance Evaluation II: Similarity Analysis}
\label{sec:CorrelationAnaly}
We measured the similarity of the extracted Twitter events against the road sensor data and scheduled sociocultural events that are parsed from the Web. The $London_2$ dataset is used for this experiment. 

The comparison results are reported in terms of classes average distance (represented with $\mu$) from their nearest authority (web) record which was described in sec.~\ref{sec:CorrGraph} along with its variance (represented by $\sigma$) and a similarity measure.  
To compute the similarities from the average distance values, we have first rescaled the averaged distances by the within-class maximum average distance (sport class distance) and then subtracted the result from one. Doing so, the similarity values are confined to be between $0$ and $1$ where closer to one values will guarantee higher similarity and values close to zero show a higher level of dissimilarity.
First row results in Table ~\ref{tab:SimTable} show different Twitter event distributions compared against the ground-truth traffic sensor data, where (latitude, longitude) polar coordinated are converted to Cartesian values~\footnote{We have used the Haversine formula explained at \url{https://en.wikipedia.org/wiki/Haversine_formula}.} and Euclidean distance used as the metric. 
\begin{table*}[ht]
\footnotesize
\caption{Similarity analysis on $London_2$ dataset: different Twitter class distributions compared against the ground-truth traffic sensor driven data distribution (first row), different Twitter class distributions compared against the ground-truth sociocultural data parsed form TimeOut London (second row).}
\resizebox{\linewidth}{!}{
\begin{tabular}{|*{9}{c|}}
\hline
&&Crime& Cultural&Food&Social&Sport&Weather&Trans. \\ \hline
\multirow{2}{*}{Road Rep.}&$\mu_D \pm \sigma$ &1.00 $\pm$ 8.79 &1.00 $\pm$ 3.67 &0.95 $\pm$ 5.14 &0.90 $\pm$ 3.74&1.69 $\pm$ 8.72 &1.72 $\pm$ 8.21&\bf{\blue{0.74 $\pm$ 2.34}} \\ \cline{2-9}
&Similarity &0.44&0.44&0.47&0.50&0.00&0.04& \bf{\blue{0.59}}\\ \hline
\multirow{2}{*}{TimeOut Rep.}&$\mu_D \pm \sigma$ &3.0 $\pm$ 31.27&\bf{\blue{2.95 $\pm$ 7.95}}&3.60 $\pm$ 45.45 &\bf{\blue{2.00 $\pm$ 7.49}}&5.26 $\pm$ 31.17 &3.77 $\pm$ 17.46&\bf{\blue{1.73 $\pm$ 5.03}} \\ \cline{2-9}
&Similarity &0.41&\bf{\blue{0.44}}&0.32&\bf{\blue{0.60}}&0.00&0.28&\bf{\blue{0.67}}\\ \hline
\end{tabular}}
\label{tab:SimTable}
\end{table*}
The results showed that the smallest average distance of $0.74$, considering the variance intervals, correspond to the Traffic events, which is a proof of an acceptable tweet classification performance. Additionally, a comparison of Twitter traffic report times with their nearest neighbour authority traffic record time-stamp, showed that $49.5\%$ of the Twitter traffic alerts are reported on average $297.5$ minutes. This is approximately $5$ hours prior to the authority's official reports. This finding highlights the advantage of utilising the social media, particularly Twitter driven knowledge in facilitating and speeding up the city traffic management and potentially smoothing the task-handling.  
Second row results in Table~\ref{tab:SimTable} show different Twitter class distributions compared with the ground-truth sociocultural data, which was parsed form Time Out London computed in the same way to traffic similarities. In our experiments, we discovered higher similarity measurements for traffic, cultural and social tweet event classes with $1.73$, $2.95$ and $2.00$ respective average distance values. This, in fact, proves that the popular and well-advertised cultural activities are potential high-traffic zones. It is important to point out that the Time Out London sociocultural event set does not separate the social events, such as special events held in pubs and restaurants, from cultural events (\ie exhibitions and ceremonies) while our proposed pipeline labels these events differently.

 \subsection{Case Studies and Web Interface}
 \label{sec:Real-time Event Annotation Temporal Statistics}

 Table~\ref{tab:CaseStudy_London2} represents case studies for $London_2$ classified events. Investigating the content of the misclassified tweets (shown in the last three rows in the table), one can spot the counter-effect of conversational language complexity on classification task where people occasionally use metaphors to emphasise their concepts \ie \textit{shooting ourselves in the foot} to describe the extend of a spoiled situation and \textit{the dessert miss the rain} to describe a lingering missing sensation. 
 \begin{table}[ht]
     \centering
     \footnotesize
     \begin{tabular}{|p{.4\textwidth}|l|c|c|c|}
     \hline
          Tweet & Location (extracted/geo-tag) & Time & Event type & Impact \\\hline\hline
          Wind 5 km/h NNW. Barometer 1012.1 mb Rising slowly. Temperature 4.9 Â°C. Rain today 0.0 mm. Humidity 73\% &(51.34,-0.08)&00:00:40& Weather & $3*0.75$ $=2.25$\\ \hline
          If Leicester win the league I will shave what's left of my hair off&(51.44,-0.03)&00:00:48& Sport & $6*0.51$ $=3.06$\\\hline
          RT ...: "When a woman and 2 children are killed it's not a domestic ""incident"" it's a crime @..."&(51.46,0.11)&00:01:02& Crime & $4*0.77$ $=3.08$\\\hline
          Left Selhurst Park at 10pm left Clapham Junction at 12.40am and got to catch a bus at Basingstoke as line closed! \#3points \#2amHome&(51.47,-0.17)&00:03:13&Transport &$7*0.93$ $=6.51$\\\hline
          Cinema at its most gripping journalism at its most courageous. Well done @... hopefully many more awards to come!&(51.29,-0.51)&00:27:20&Cultural& $1*0.51$ $=0.51$ \\\hline
          I've seen way too many horror films in my time to not feel at ease wandering around uni and/or halls in the windy darkness&(51.44,-0.03)&01:08:22&Weather& $1*0.86$ $=0.86$\\\hline
          ... im legit so hyped for the Marina concert&(51.52,-0.02)& 01:09:26&Cultural &$1*0.89$ $=0.89$\\\hline
          Traffic is disgusting this morning ðŸ˜£ ðŸ”«&(51.51,0.06)&07:51:15&Transport&$2*0.82$ $=1.64$\\\hline
          Good morning campus @ King's College London &(51.51, -0.11)& 07:51:09& Social&$1*0.98$ $=0.98$\\\hline
          Man's body found after triple murder @SkyNews&(51.39, 0.23)& 07:54:17& Crime&$2*0.61$ $=1.22$\\\hline
          RT @...: @... please attend debate consider voting for bigger spend leads to better outcomes for health \&amp; economy&(51.64, -0.38)&07:59:17& Social &$2*0.69$ $=1.38$\\\hline
          @... Ayew's a villa player isn't he? That's us shooting ourselves in the foot.&(51.51,-0.21)&00:03:05&Crime& $1*0.92$ $=0.92$\\\hline
          And i miss you the desserts miss the rain....&(51.49, -0.35)& 00:03:58& Weather&$2*0.78$ $=1.56$\\\hline
          @... over 8 shots saved by keeper or cleared off the line. Could've been 5-0. These games happen - just have to keep going. \#OurYear?&(51.51, -0.21)&00:31:09&Transport&$1*0.92$ $=0.92$\\\hline
            \end{tabular}
     \caption{Case studies for $London_2$ Twitter data classified events}
     \label{tab:CaseStudy_London2}
 \end{table}
 
\begin{figure}[ht]
\begin{center}
\includegraphics[width=0.8\columnwidth]{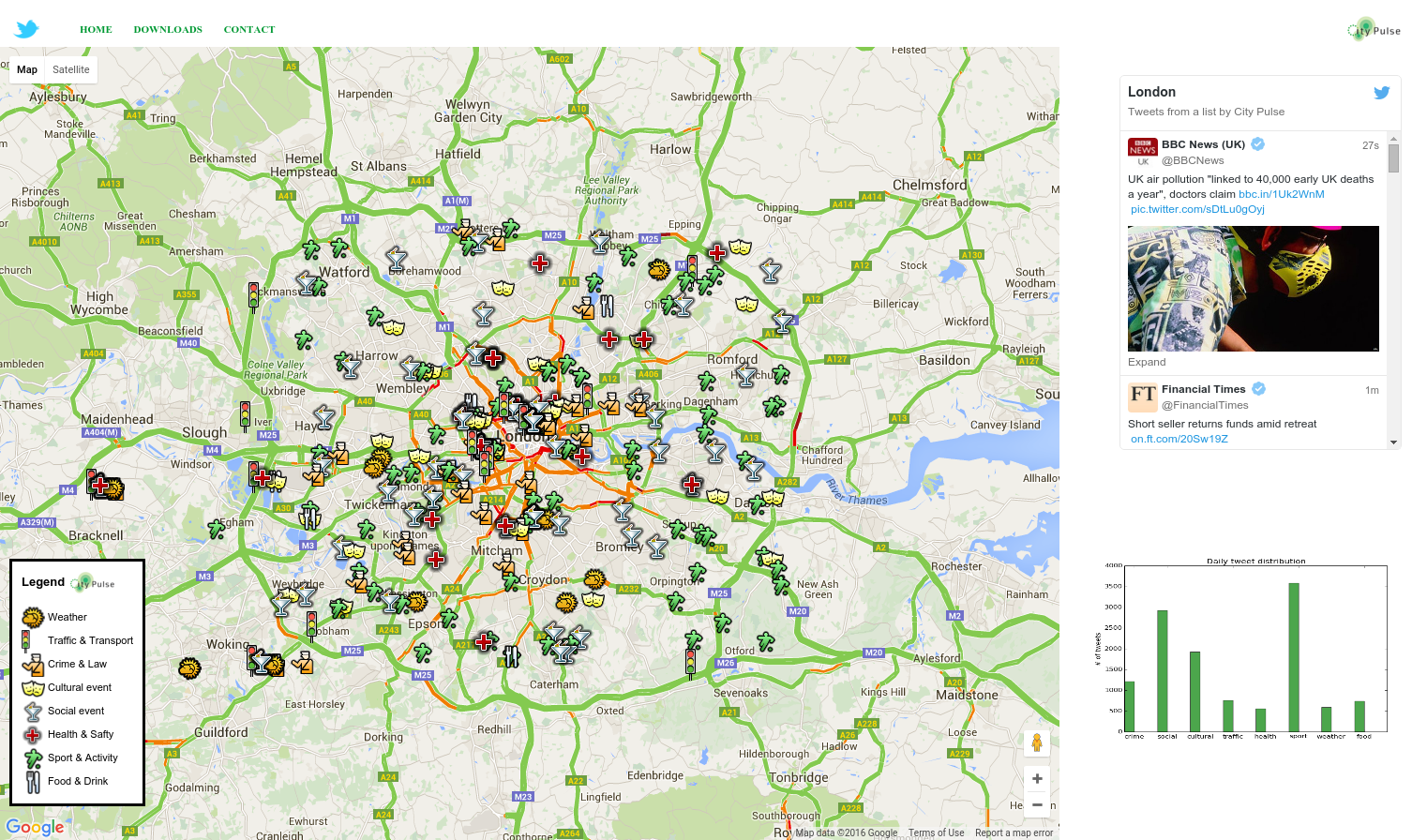}
\end{center}
\caption{A screen shot of the developed web-interface}
\label{fig:Map-screenshot}
\end{figure}

 While in a recent study Zuao \etal~\cite{DBLP:conf/acl/ZhouGH16} proposed to model the intrinsic geometry of tweets through a low-rank, non-linear manifold to visualize the tweets distribution in a two dimensional Euclidean space, in this study we have focused on a visualisation in the real space. 
 To facilitate this, we have developed a Web interface. The interface displays the classification results on a Google map in near real-time and it is composed of four elements; $i)$ Google map canvas layer on which the processed and annotated tweets are displayed with their class-identical icons; $ii)$ a live London traffic layer from Google traffic API - code coloured paths on the map; $iii)$ a bar chart panel which presents the class distribution histogram of daily tweets; and $iv)$ a panel for displaying Twitter time line.
The map data is being updated every 60 seconds by appending the past minute's tweets to existing ones up to a 60-minutes time window. In practice, the whole data will be updated on hourly basis. Clicking on each event a dialogue box is shown on the map which reveals the underlying tweet content along with its time-stamp. The twitter user id and the names are anonymised for privacy purpose.
The web interface~\footnote{The interface is accessible at http://iot.ee.surrey.ac.uk/citypulse-social/.} utilises javascript and HTML coding to read the data results saved in a CSV data structure format and displays the tweets on the map in near real time with less than 60 seconds latency.
Fig.~\ref{fig:Map-screenshot} shows an screen shot of the web-interface~\footnote{The interface is accessible at http://iot.ee.surrey.ac.uk/citypulse-social/.}.

\section{Conclusions}
\label{sec:conclusion}
In this paper, we developed a multi-label event detection framework for 
live annotation and classification of Twitter data, in a smart city context. 

We have introduced a set of common event dictionaries with minimal overlaps 
for facilitating the detection of city-related events and developed a GUI to facilitate the ground-truth annotation of tweets for machine learning tasks. 
More importantly, we have developed a novel multi-view learning framework that utilises the Convolutional Neural Network (CNN) features with CRF-dictionary driven NER tags for sentence-level event annotation. 
The proposed model is tested on geographically different English speaking cities and the results have proved promising. 
The evaluation results showed that our proposed 
solution is capable of annotating tweet words with an averaged accuracy of $81\%$ over all event classes. And the multi-view deep learning model boosted the performance over a single-view event classification approach by 5\%. 

 We have also performed similarity analysis which showed how authority driven traffic reports and scheduled sociocultural events can affect the traffic pattern and how citizens project on them through social media. The evaluations showed that 49.5\% of the Twitter traffic comments are reported approximately five hours prior to authority's official records and the scheduled sociocultural events have observed influencing the distribution of the twitter comments of traffic class, along as cultural and social classes. The study highlights the possibility of utilising social media as human probes for realising a real-time Physical-Cyber-Social platform for ranking, completing and potentially speeding up the city service deliveries. 
 Finally, we have developed a live stream analysis interface to present the analysis results and case studies on a Google map. The proposed interface enables the public to visualise their city and neighbourhood event patterns.
 
 The proposed model serves as a proof of concept and improvements can be made in several stages of the pipeline. For example, the model can be tested on non-English twitter streams, the multi-view learning labels can be updated with an online recursive learning model which is more adaptive and provides performance feedback to the rest of the pipeline.



\textbf{Acknowledgement:} 
This work has been carried out in the scope of the European Commission's Seventh Framework Programme funded project CityPulse (FP7-609035).

\footnotesize{
\bibliographystyle{plain}
\bibliography{paper.bbl}

\begin{thebibliography}{10}

\bibitem{Adelson85spatiotemporalenergy}
Edward~H. Adelson and James~R. Bergen.
\newblock Spatiotemporal energy models for the perception of motion.
\newblock {\em J. OPT. SOC. AM. A}, 2(2):284--299, 1985.

\bibitem{LingPipe}
Alias-i.
\newblock Lingpipe 4.1.0.
\newblock \url{http://alias-i.com/lingpipe}.
\newblock Accessed: 2015-05-05.

\bibitem{Pramod-ACM-2015}
Pramod Anantharam, Payam Barnaghi, Krishnaprasad Thirunarayan, and Amit~P.
  Sheth.
\newblock Extracting city traffic events from social streams.
\newblock In {\em {ACM} Transactions on Intelligent Systems and Technology},
  volume~-, New York, NY, USA, 2015. ACM.

\bibitem{Asratian:1998:BGA:294028}
Armen~S. Asratian, Tristan M.~J. Denley, and Roland H\"{a}ggkvist.
\newblock {\em Bipartite Graphs and Their Applications}.
\newblock Cambridge University Press, New York, NY, USA, 1998.

\bibitem{DBLP:conf/icwsm/BeckerNG11}
Hila Becker, Mor Naaman, and Luis Gravano.
\newblock Beyond trending topics: Real-world event identification on twitter.
\newblock In Lada~A. Adamic, Ricardo~A. Baeza{-}Yates, and Scott Counts,
  editors, {\em Proceedings of the Fifth International Conference on Weblogs
  and Social Media, Barcelona, Catalonia, Spain, July 17-21, 2011}. The {AAAI}
  Press, 2011.

\bibitem{Blissent:2010}
Jennifer B\'{e}lissent.
\newblock Getting clever about smart cities: new opportunities require new
  business models.
\newblock In {\em Vendor Strategy Professionals}, 2010.

\bibitem{Blissent:2013}
Jennifer B\'{e}lissent and Frederic Giron.
\newblock Service providers accelerate smart city projects.
\newblock In {\em Forrester}, July 2013.

\bibitem{Bengio:2003:NPL:944919.944966}
Yoshua Bengio, R{\'e}jean Ducharme, Pascal Vincent, and Christian Janvin.
\newblock A neural probabilistic language model.
\newblock {\em J. Mach. Learn. Res.}, 3:1137--1155, March 2003.

\bibitem{SmartCities2011}
Anthony Bernal and Chief Programmer.
\newblock Building a smarter planet, one city at a time.
\newblock online resource, May 2011.
\newblock Industry Solutions, IBM.

\bibitem{Burke06participatorysensing}
J.~Burke, D.~Estrin, M.~Hansen, A.~Parker, N.~Ramanathan, S.~Reddy, and M.~B.
  Srivastava.
\newblock Participatory sensing.
\newblock In {\em In: Workshop on World-Sensor-Web (WSW’06): Mobile Device
  Centric Sensor Networks and Applications}, pages 117--134, 2006.

\bibitem{conf/nips/ChenZX10}
Ning Chen, Jun Zhu, and Eric~P. Xing.
\newblock Predictive subspace learning for multi-view data: a large margin
  approach.
\newblock In John~D. Lafferty, Christopher K.~I. Williams, John Shawe-Taylor,
  Richard~S. Zemel, and Aron Culotta, editors, {\em NIPS}, pages 361--369.
  Curran Associates, Inc., 2010.

\bibitem{Collobert:2011:NLP:1953048.2078186}
Ronan Collobert, Jason Weston, L{\'e}on Bottou, Michael Karlen, Koray
  Kavukcuoglu, and Pavel Kuksa.
\newblock Natural language processing (almost) from scratch.
\newblock {\em J. Mach. Learn. Res.}, 12:2493--2537, November 2011.

\bibitem{C14-1008}
Cicero dos Santos and Maira Gatti.
\newblock Deep convolutional neural networks for sentiment analysis of short
  texts.
\newblock In {\em Proceedings of COLING 2014, the 25th International Conference
  on Computational Linguistics: Technical Papers}, pages 69--78. Dublin City
  University and Association for Computational Linguistics, 2014.

\bibitem{DBLP:conf/wf-iot/FarajiDavarKB16}
Nazli FarajiDavar, Sefki Kolozali, and Payam~M. Barnaghi.
\newblock Physical-cyber-social similarity analysis in smart cities.
\newblock In {\em 3rd {IEEE} World Forum on Internet of Things, WF-IoT 2016,
  Reston, VA, USA, December 12-14, 2016}, pages 484--489, 2016.

\bibitem{Filipponi:2010}
Luca Filipponi, Andrea Vitaletti, Giada Landi, Vincenzo Memeo, Giorgio Laura,
  and Paolo Pucci.
\newblock Smart city: An event driven architecture for monitoring public spaces
  with heterogeneous sensors.
\newblock In {\em Fourth International Conference in Sensor Technologies and
  Applications SENSORCOMM}, pages 281--286. IEEE, 2010.

\bibitem{conf/ciarp/FischerI12}
Asja Fischer and Christian Igel.
\newblock An introduction to restricted boltzmann machines.
\newblock In Luis Álvarez, Marta Mejail, Luís~Gómez Déniz, and Julio~C.
  Jacobo, editors, {\em CIARP}, volume 7441 of {\em Lecture Notes in Computer
  Science}, pages 14--36. Springer, 2012.

\bibitem{Fleet85EnergyModel}
D.~Fleet, H.~Wagner, and D.~Heeger.
\newblock Nueral encoding of the binocular disparity: Energu models, position
  shifts and phase shift.
\newblock {\em Vision Research}, 36(12):1839--1857, 1996.

\bibitem{Glorot+al-ICML-2011}
Xavier Glorot, Antoine Bordes, and Yoshua Bengio.
\newblock Domain adaptation for large-scale sentiment classification: A deep
  learning approach.
\newblock In {\em Proceedings of theTwenty-eight International Conference on
  Machine Learning (ICML'11)}, volume~27, pages 97--110, June 2011.

\bibitem{Grishman:2002:REE:1289189.1289229}
Ralph Grishman, Silja Huttunen, and Roman Yangarber.
\newblock Real-time event extraction for infectious disease outbreaks.
\newblock In {\em Proceedings of the Second International Conference on Human
  Language Technology Research}, HLT '02, pages 366--369, San Francisco, CA,
  USA, 2002. Morgan Kaufmann Publishers Inc.

\bibitem{Haklay:2008:OUS:1477057.1477249}
Mordechai~(Muki) Haklay and Patrick Weber.
\newblock Openstreetmap: User-generated street maps.
\newblock {\em IEEE Pervasive Computing}, 7(4):12--18, October 2008.

\bibitem{Hinton:2006:FLA:1161603.1161605}
Geoffrey~E. Hinton, Simon Osindero, and Yee-Whye Teh.
\newblock A fast learning algorithm for deep belief nets.
\newblock {\em Neural Comput.}, 18(7):1527--1554, July 2006.

\bibitem{Journal/geo/Karney13}
Charles F.~F. Karney.
\newblock Algorithms for geodesics.
\newblock In {\em Journal of Geodesy}, volume~87, pages 43--55. Springer, 2013.

\bibitem{kehoe:Cosgrove:2011}
Michael Kehoe, Michael Cosgrove, SD~Gennaro, Colin Harrison, Wim Harthoorn,
  John Hogan, Pam~Nesbitt John~Meegan, and Christina Peters.
\newblock Smarter cities series: a foundation for understanding ibm smarter
  cities.
\newblock In {\em An IBM Redguide publication}. An IBM Redguide publication,
  2011.

\bibitem{Koller+Friedman:09}
D.~Koller and N.~Friedman.
\newblock {\em Probabilistic Graphical Models: Principles and Techniques}.
\newblock MIT Press, 2009.

\bibitem{daume11spectral}
Abhishek Kumar, Piyush Rai, and Hal {Daum\'e III}.
\newblock Co-regularized multi-view spectral clustering.
\newblock In {\em Proceedings of the Conference on Neural Information
  Processing Systems (NIPS)}, Granada, Spain, 2011.

\bibitem{TwitterDataAnalytics2013}
Shamanth Kumar, Fred Morstatter, and Huan Liu.
\newblock {\em Twitter Data Analytics}.
\newblock Springer, New York, NY, USA, 2013.

\bibitem{Lampos:2012:NES:2337542.2337557}
Vasileios Lampos and Nello Cristianini.
\newblock Nowcasting events from the social web with statistical learning.
\newblock {\em ACM Trans. Intell. Syst. Technol.}, 3(4):72:1--72:22, September
  2012.

\bibitem{Lindsey2010}
Greg Lindsay.
\newblock Cisco’s big bet on new songdo: creating cities from scratch, 2010.
\newblock http://www.fastcompany.com/.

\bibitem{DBLP:conf/ideal/LiuLXCY08}
Mingrong Liu, Yicen Liu, Liang Xiang, Xing Chen, and Qing Yang.
\newblock Extracting key entities and significant events from online daily
  news.
\newblock In {\em Intelligent Data Engineering and Automated Learning - {IDEAL}
  2008, 9th International Conference, Daejeon, South Korea, November 2-5, 2008,
  Proceedings}, pages 201--209, 2008.

\bibitem{Marquez:2008:SRL:1403157.1403158}
Llu\'{\i}s M\`{a}rquez, Xavier Carreras, Kenneth~C. Litkowski, and Suzanne
  Stevenson.
\newblock Semantic role labeling: An introduction to the special issue.
\newblock {\em Comput. Linguist.}, 34(2):145--159, June 2008.

\bibitem{DBLP:journals/corr/abs-1206-4609}
Roland Memisevic.
\newblock On multi-view feature learning.
\newblock {\em CoRR}, abs/1206.4609, 2012.

\bibitem{Moraru:2012}
Dunja Mladeni\'{c} and Alexandra Moraru.
\newblock Complex event processing and data mining for smart cities.
\newblock In {\em Conference on Data Mining and Data Warehouses (SiKDD 2012)},
  2012.

\bibitem{Naphade:2011:SCI:1999160.1999174}
Milind Naphade, Guruduth Banavar, Colin Harrison, Jurij Paraszczak, and Robert
  Morris.
\newblock Smarter cities and their innovation challenges.
\newblock {\em Computer}, 44(6):32--39, June 2011.

\bibitem{conf/airs/OkamotoK09}
Masayuki Okamoto and Masaaki Kikuchi.
\newblock Discovering volatile events in your neighborhood: Local-area topic
  extraction from blog entries.
\newblock In Gary~Geunbae Lee, Dawei Song, Chin-Yew Lin, Akiko~N. Aizawa,
  Kazuko Kuriyama, Masaharu Yoshioka, and Tetsuya Sakai, editors, {\em AIRS},
  volume 5839 of {\em Lecture Notes in Computer Science}, pages 181--192.
  Springer, 2009.

\bibitem{kernel-based-Learning2013}
Novi Quadrianto and Christoph~H Lampert.
\newblock {\em Kernel-based learning}.
\newblock In: Encyclopedia of systems biology. Springer New York, New York,
  2013.

\bibitem{Ritter:2012:ODE:2339530.2339704}
Alan Ritter, Mausam, Oren Etzioni, and Sam Clark.
\newblock Open domain event extraction from twitter.
\newblock In {\em Proceedings of the 18th ACM SIGKDD International Conference
  on Knowledge Discovery and Data Mining}, KDD '12, pages 1104--1112, New York,
  NY, USA, 2012. ACM.

\bibitem{Sasaki:2008:MMN:1572306.1572318}
Yutaka Sasaki, Yoshimasa Tsuruoka, John McNaught, and Sophia Ananiadou.
\newblock How to make the most of ne dictionaries in statistical ner.
\newblock In {\em Proceedings of the Workshop on Current Trends in Biomedical
  Natural Language Processing}, BioNLP '08, pages 63--70, Stroudsburg, PA, USA,
  2008. Association for Computational Linguistics.

\bibitem{Sheth2009}
Amit~P. Sheth.
\newblock Citizen sensing, social signals, and enriching human experience.
\newblock In {\em {IEEE} Transactions on Internet Computing}, volume~13, pages
  87--92, 2015.

\bibitem{conf/nldb/TanevPA08}
Hristo Tanev, Jakub Piskorski, and Martin Atkinson.
\newblock Real-time news event extraction for global crisis monitoring.
\newblock In Epaminondas Kapetanios, Vijayan Sugumaran, and Myra Spiliopoulou,
  editors, {\em NLDB}, volume 5039 of {\em Lecture Notes in Computer Science},
  pages 207--218. Springer, 2008.

\bibitem{S14-2033}
Duyu Tang, Furu Wei, Bing Qin, Ting Liu, and Ming Zhou.
\newblock Coooolll: A deep learning system for twitter sentiment
  classification.
\newblock In {\em Proceedings of the 8th International Workshop on Semantic
  Evaluation (SemEval 2014)}, pages 208--212. Association for Computational
  Linguistics, 2014.

\bibitem{wang2012automatic}
Xiaofeng Wang, Matthew~S Gerber, and Donald~E Brown.
\newblock Automatic crime prediction using events extracted from twitter posts.
\newblock In {\em Social Computing, Behavioral-Cultural Modeling and
  Prediction}, pages 231--238. Springer, 2012.

\bibitem{StatisticalMethods4Immunogenicity2015}
Harry Yang, Jianchun Zhang, Binbing Yu, and Wei Zhao.
\newblock {\em Statistical Methods for Immunogenicity Assessment}.
\newblock Chapman and Hall/CRC, Sep 2015.

\bibitem{DBLP:conf/acl/ZhouGH16}
Deyu Zhou, Tianmeng Gao, and Yulan He.
\newblock Jointly event extraction and visualization on twitter via
  probabilistic modelling.
\newblock In {\em Proceedings of the 54th Annual Meeting of the Association for
  Computational Linguistics, {ACL} 2016, August 7-12, 2016, Berlin, Germany,
  Volume 1: Long Papers}, 2016.

\bibitem{PCS/Zhou16}
Yuchao Zhou, Suparna De., and Klaus Moessner.
\newblock Real world city event extraction from twitter data streams.
\newblock {\em Procedia Computer Science}, 98:443--448, 2016.

\end{thebibliography}
      }
\end{document}